\documentclass[fleqn,usenatbib]{mnras}

\usepackage{newtxtext,newtxmath}
\usepackage[T1]{fontenc}
\usepackage{ae,aecompl}

\usepackage{graphicx}	
\usepackage{amsmath}	
\usepackage{amssymb}	

\usepackage[caption = false]{subfig}
\usepackage{arydshln}
\usepackage{CJKutf8}

\graphicspath{{plots/}}

\newcommand{\Mpc}{~h^{-1}~{\rm Mpc}}
\newcommand{\Kpc}{~h^{-1}~{\rm kpc}}
\newcommand{\Gpc}{~h^{-1}~{\rm Gpc}}

\newcommand{\Fig}[1]{Fig.~\ref{#1}}

\newcommand{\hMpc}{{\ifmmode{~h^{-1}{\rm Mpc}}\else{$h^{-1}$Mpc}\fi}}
\newcommand{\hkpc}{{\ifmmode{~h^{-1}{\rm kpc}}\else{$h^{-1}$kpc}\fi}}
\newcommand{\hGpc}{{\ifmmode{~h^{-1}{\rm Gpc}}\else{$h^{-1}$Gpc}\fi}}
\newcommand{\hMsun}{{\ifmmode{~h^{-1}{\rm {M_{\odot}}}}\else{$h^{-1}{\rm{M_{\odot}}}$}\fi}}
\newcommand{\Mstar}{{\ifmmode{~M_{*}}\else{$M_{*}$}\fi}}
\newcommand{\Mhalo}{{\ifmmode{~M_{\rm halo}}\else{$M_{\rm halo}$}\fi}}
\newcommand{\ltsima}{$\; \buildrel < \over \sim \;$}
\newcommand{\gtsima}{$\; \buildrel > \over \sim \;$}
\newcommand{\lsim}{\lower.5ex\hbox{\ltsima}}
\newcommand{\gsim}{\lower.5ex\hbox{\gtsima}}

\newcommand{\ahf}{\textsc{AHF}}

\newcommand{\gadgetx}{\textsc{Gadget-X}}
\newcommand{\gadgetmusic}{\textsc{Gadget-MUSIC}}
\newcommand{\galacticus}{\textsc{Galacticus}}
\newcommand{\sag}{\textsc{SAG}}
\newcommand{\sage}{\textsc{SAGE}}

\title[The Three Hundred Project]{The Three Hundred Project: A large catalogue of theoretically modelled galaxy clusters for cosmological and astrophysical applications.}

\author[Cui et al.]{\parbox{\textwidth}{
Weiguang Cui,$^{1}$\thanks{E-mail: weiguang.cui@uam.es}
Alexander Knebe,$^{1,2,3}$\thanks{alexander.knebe@uam.es} Gustavo Yepes,$^{1,2}$\thanks{gustavo.yepes@uam.es} Frazer Pearce,$^{4}$
Chris Power,$^{3}$ Romeel Dave,$^{5}$ Alexander Arth,$^{6,7}$ Stefano Borgani,$^{8,9,10}$ Klaus Dolag,$^{6,11}$ 
Pascal Elahi,$^{3}$ Robert Mostoghiu,$^{1}$ Giuseppe Murante,$^{9}$ Elena Rasia,$^{9}$ Doris Stoppacher,$^{1,12}$ 
Jesus Vega-Ferrero,$^{13}$ Yang Wang,$^{14}$ Xiaohu Yang,$^{15,16}$ Andrew Benson,$^{17}$ Sof\'ia A. Cora,$^{18,19}$ 
Darren J. Croton,$^{20}$ Manodeep Sinha,$^{20}$ Adam R. H. Stevens,$^{3}$ Cristian A. Vega-Mart\'inez,$^{18}$
Jake Arthur,$^{4}$ Anna S. Baldi,$^{21,22}$ Rodrigo Ca\~nas,$^{3}$ Giammarco Cialone,$^{21}$
Daniel Cunnama,$^{23,24}$ Marco De Petris,$^{21,25}$ Giacomo Durando,$^{26}$ Stefano Ettori,$^{27,28}$
Stefan Gottl\"{o}ber,$^{29}$ Sebasti\'an~E.~Nuza,$^{30,31}$ Lyndsay J. Old,$^{32}$ 
Sergey Pilipenko,$^{33}$ Jenny G. Sorce,$^{34,29}$ Charlotte Welker$^{3,35}$\newline
\emph{\normalsize Affiliations are listed at the end of the paper}
}}

\date{Accepted XXX. Received YYY; in original form ZZZ}
\pubyear{2018}

\begin{document}
\label{firstpage}
\pagerange{\pageref{firstpage}--\pageref{lastpage}}
\maketitle

\begin{abstract}
We introduce the {\sc The Three Hundred} project, an endeavour to model 324 large galaxy clusters with full-physics hydrodynamical re-simulations. Here we present the dataset and study the differences to observations for fundamental galaxy cluster properties and scaling relations. We find that the modelled galaxy clusters are generally in reasonable agreement with observations with respect to baryonic fractions and gas scaling relations at redshift $z=0$. However, there are still some (model-dependent) differences, such as central galaxies being too massive, and galaxy colours ($g-r$) being bluer (about 0.2 dex lower at the peak position) than in observations. The agreement in gas scaling relations down to $10^{13}\hMsun$ between the simulations indicates that particulars of the sub-grid modelling of the baryonic physics only has a weak influence on these relations. We also include -- where appropriate -- a comparison to three semi-analytical galaxy formation models as applied to the same underlying dark matter only simulation. 
All simulations and derived data products are publicly available.
\end{abstract}



\section{Introduction}

Galaxy clusters are the largest gravitationally bound objects in the Universe and as such they provide a host environment for testing both cosmology models and theories of galaxy evolution. Their formation depends both on the underlying cosmological framework and the details of the baryonic physics that is responsible for powerful feedback processes. Amongst others, these mechanisms regulate the observed properties of the Intra-Cluster Medium (ICM), the size of the central brightest cluster galaxy and the number and properties of the satellite galaxies orbiting within a common dark matter envelope. Clusters of galaxies can therefore be considered to be large cosmological laboratories that are useful for pinning down both cosmological parameters and empirical models of astrophysical processes acting across a range of coupled scales.

Concerted effort, from both observational and theoretical perspectives, has been devoted to improve our understanding of the formation and evolution of galaxy clusters. On the observational side, multi-wavelength telescopes are designed to observe different properties of galaxy clusters: radio and far infrared data provide information on the cold gas; optical data focusses attention on the stellar properties and provides input to gravitational lensing analyses which target the dark-matter (DM) component; millimetre and X-ray observations target the ICM. 
In parallel with these observational programmes, hydrodynamical simulations of the formation and evolution of galaxy clusters have been a very powerful tool to interpret and guide observations for more than 20 years \citep{Evrard1996,Bryan1998}. However, these extremely large objects with masses $M\geq 10^{15}\hMsun$ are very rare and can only be found in large volumes $V\gg (100\hMpc)^3$. But modelling such volumes with all the relevant dark matter \textit{and} baryonic physics, while obtaining sufficient mass and spatial resolution at the same time, is challenging. Therefore, the most commonly used approach is to perform so-called `zoom' simulations, i.e. selecting an object of interest from a parent dark matter simulation and only adding baryonic physics (at a much higher resolution) in a region about that object. This strategy has led to valuable results, but in order to be of statistical significance one would need to run hundreds -- if not thousands -- of such zoom simulations, which is what workers in the field are striving for at the moment.

Recent years have seen great advances in the direction of generating substantial samples of highly resolved galaxy cluster simulations that include all the relevant baryonic processes, e.g. the 500 `MUSIC' clusters \citep{MUSICI}, the sample of 29 clusters of \citet{Planelles2013}, the 10 `Rhapsody-G' clusters \citep{Wu2015}, the 390 `MACSIS' clusters \citep{Barnes2017a}, the 30 `Cluster-EAGLE' \citep{Barnes2017b} and 24 related `Hydrangea' clusters \citep{Bahe2017}. The mass resolution of these zoom simulations varies from sample to sample covering the range of dark matter particle masses $m_{\rm DM}=9.7\times 10^{6}\hMsun$ for `Hydrangea' and `Cluster-EAGLE' up to $4.4\times 10^{9}\hMsun$ for the large `MACSIS' sample. There are additionally cluster samples extracted from full box simulations, e.g. `cosmo-OWLS' \citep{LeBrun2014} and its follow-up `BAHAMAS' \citep{McCarthy2017} featuring hundreds of galaxy clusters, but the majority with masses lower than $10^{14.5}\hMsun$ and at a mass resolution of $m_{\rm DM}\sim 4 \times 10^{9}\hMsun$.


In a series of precursor papers \citep[i.e. the `nIFTy cluster comparison project' introduced in][]{Sembolini2016, Sembolini2016a} we investigated the differences in cluster properties arising from simulating \textit{one} individual galaxy cluster with a variety of different numerical techniques including standard SPH, modern\footnote{We define `modern' as those SPH implementations that adopt an improved treatment of discontinuities.} SPH, and (moving) mesh codes. The results obtained there led us to the choice of using the modern SPH code \textsc{GADGET-X} which includes an improved SPH scheme and the implementation of black hole (BH) and active galactic nuclei (AGN) feedback compared to our fiducial \textsc{GADGET-MUSIC} code.

The primary goal of this paper is to introduce {\sc The Three Hundred} project and its associated data set\footnote{The data (ca. 50 TB of simulation data and 4TB of halo catalogues) are stored on a server to which access will be granted upon request to either AK or GY.} that maximizes the ratio between number of objects and mass resolution: 324 regions of radius $15\hMpc$ -- having a cluster with mass $M_{200}>6.42\times 10^{14}\hMsun$ at its centre -- have been modelled with a combined mass resolution of $m_{\rm DM} + m_{\rm gas} = 1.5\times 10^{9}\hMsun$. This is, in fact, the same resolution as used for our previous `MUSIC' clusters, but the difference here lies in an improved modelling of sub-grid physics and an application of a modern numerical Smooth-Particle-Hydrodynamics (SPH) scheme. We detail the hydro-simulations, and the procedures for producing the cluster catalogue. We also present generic results, such as the dynamical state, baryon fraction, and optical/gas scaling relations. In addition, we add to the plots -- where possible -- the results from three semi-analytical galaxy formation models \galacticus, \sag, and \sage, noting that they have been applied to the same dark-matter-only simulation that formed the basis for the selection of the clusters presented here \citep[see][for the public release of the corresponding catalogues]{Knebe2018}. Although this is not the first time that a joint analysis of hydrodynamical simulations with SAMs has been performed \citep[for example,][]{Saro2010, Cui2011, Monaco2014, Guo2016}, it is, to our knowledge, the first time such an approach has been applied to a large number of galaxy clusters. Detailed comparisons between the models and further investigation into different aspects of the cluster properties will be addressed in following companion papers.

The paper is structured as follows: we begin by describing the properties of the cluster sample in section \ref{sec:cluster}, which also includes a description of the hydrodynamical methods and of the semi-analytic models. We briefly present our results for cluster bulk properties in section \ref{sec:bulk}, and for the relevant relations in different wavebands in section \ref{sec:sr}. We conclude our results in \ref{sec:conclusion}.

\section{The galaxy cluster sample} \label{sec:cluster}

The basis of our dataset has been formed by extracting 324 spherical regions centred on each of the most massive clusters identified at $z=0$ by the \textsc{Rockstar}\footnote{\url{https://bitbucket.org/gfcstanford/rockstar}} halo finder \citep{Behroozi2013} within the dark-matter-only MDPL2, MultiDark simulation \citep{Klypin2016}.\footnote{The MultiDark simulations are publicly available at the \url{https://www.cosmosim.org} database.} The MDPL2 simulation utilises the cosmological parameters shown in Table \ref{table:cosmo} which are those of the {\it Planck} mission \citep{Planck2016}. The MDPL2 is a periodic cube of comoving length $1 \Gpc$ containing $3840^3$ dark matter particles, each of mass $1.5 \times 10^9 \hMsun$.

\begin{table}
  \caption{Parameters of {\sc The Three Hundred} simulations}
  \label{table:cosmo}
  \begin{tabular}{lll}
    \hline
    & Value & Description\\
    \hline
    $\Omega_{\rm M}$ & 0.307 & Total matter density parameter\\
    $\Omega_{\rm B}$ & 0.048 & Baryon density parameter\\
    $\Omega_{\rm \Lambda}$ & 0.693 & Cosmological constant density parameter\\
    $h$ & 0.678  & Hubble constant in units of 100 km/s/Mpc\\
    $\sigma_8$ & 0.823 & Power spectrum normalization\\
    $n_{\rm s}$ & 0.96  & Power index\\
    $z_{\rm init}$ & 120 & Initial redshift \\
    $\epsilon_{\rm phys}$ & 6.5 & Plummer equivalent softening in \hkpc\\
    $L$ & 1  & Size of the MDPL2 simulation box in \hGpc\\
    $R_{\rm resim}$ & 15 & Radius for each re-simulation region in \hMpc\\
    $M_{\rm DM}$ & 12.7  & dark matter particle mass in $10^8$ \hMsun \\
    $M_{\rm gas}$ & 2.36 & gas particle mass in $10^8$ \hMsun \\
    \hline
  \end{tabular}
\end{table}

\subsection{The full-physics hydrodynamical simulations} \label{sec:simulations}
\label{sec:match}

The 324 clusters at the centre of each re-simulation region were selected initially as those with the largest halo virial mass\footnote{The halo virial mass is defined as the mass enclosed inside an overdensity of $\sim$98 times the critical density of the Universe \citep{Bryan1998}.} at $z=0$ with $M_{\rm vir} \gtrapprox 8 \times 10^{14} \hMsun$. The centres of their dark matter haloes serve as the centre of a spherical region with radius 15$\Mpc$, for which initial conditions with multiple levels of mass refinement have been generated using the fully parallel \textsc{Ginnungagap}\footnote{\url{https://github.com/ginnungagapgroup/ginnungagap}} code. Dark matter particles within the highest resolution Lagrangian regions are split into dark matter and gas particles, according to the assumed cosmological baryon fraction listed in Table \ref{table:cosmo}. Our mass resolution is a factor of three better than that used for the 390 `MACSIS' clusters \citep{Barnes2017a}. We further highlight that our re-simulation regions have the same mass resolution as the original dark matter only simulation upon which the SAMs are based. The dark matter particles outside this region are successively degraded in multiple layers (with a shell thickness of $\sim 4 \Mpc$) with lower mass resolution particles (increased by 8 times for each layer) that eventually provide the same tidal fields yet at a much lower computational costs than in the original simulation\footnote{The initial conditions for these clusters are publicly available in \textsc{Gadget} format and can be downloaded from \url{ http://music.ft.uam.es} upon request. We have also produced higher resolution initial conditions corresponding to an equivalent resolution of $7680^3$ particles, for a sub-sample of the cluster catalogue.}. The size of the re-simulated region is much larger than the virial radius of the cluster it surrounds. As such, each region also contains many additional groups and filamentary structure which may or may not be physically associated with the cluster they surround.

The initial conditions -- also publicly available -- were run with the `modern' SPH code \gadgetx\ and snapshots of the simulations stored for a set of pre-selected redshifts. A total of 128 different snapshots have been stored for each simulation from redshift $z = $ 17 to 0. We also ran the same simulations with our fiducial \gadgetmusic\ code \citep{MUSICI}. Both codes are based on the gravity solver of the {\sc GADGET3} Tree-PM code (an updated version of the {\sc GADGET2} code; \citealt{Springel2005}). While both use smooth-particle hydrodynamics (SPH) to follow the evolution of the gas component, they apply different SPH techniques as well as rather distinct models for the sub-resolution physics. \gadgetx\ includes an improved SPH scheme \citep{Beck2016} with artificial thermal diffusion, time-dependent artificial viscosity, high-order Wendland C4 interpolating kernel and wake-up scheme. These improvements advance the SPH capability of following gas-dynamical instabilities and mixing processes by better describing the discontinuities and reducing the clumpiness instability of gas. They also minimize the viscosity away from shock regions and especially in rotating shears. \gadgetmusic\ uses the classic entropy-conserving SPH formulation with a 40 neighbour M3 interpolation kernel. The differences in baryon treatment have been summarized in Table~\ref{table:bm}. For more details and the implications of the code differences we refer the reader to our comparison papers \citep[][]{Sembolini2016,Sembolini2016a}.

\begin{table*}
  \caption{Baryonic models for the two simulation codes.}
  \label{table:bm}
  \begin{tabular}{lll}
    \hline
    Baryon physics & \gadgetmusic\ & \gadgetx\ \\
    \hline
      Gas treatment	&  \\
     \hdashline
    homogeneous UV background	& \cite{Haardt2001} & \cite{Haardt1996} \\
	Cooling		&	metal independent &	metal dependent \citep{Wiersma2009} \\
    \hline
      Star formation and stellar feedback &	\\
      \hdashline
    Stellar model	&	\cite{Springel2003}	&	\cite{Tornatore2007}\\
    Threshold for star forming & 0.1 cm$^{-3}$  &  0.1 cm$^{-3}$ \\
	IMF				&	\cite{Salpeter1955}	&	\cite{Chabrier2003} \\
	Kinetic	feedback& \cite{Springel2003}	&	\cite{Springel2003}	\\
    Wind velocity	& 400 km/s			&	350 km/s\\
	Thermal feedback& 2-phase model \citep{1997MNRAS.284..235Y} &	only set the hot phase temperature \\
    Gas mass loss	& via galactic winds	& no \\
    \hline
     BH and AGN feedback	&  \\
     \hdashline
    BH seeding & no  &  $M_{\rm bh} = 5\times 10^6 \hMsun$ for $M_{\rm FoF} \geq 2.5\times 10^{11} \hMsun$\\
    BH growth & no	& Individual accretion of hot and cold gas \\
    AGN feedback & no & \cite{Steinborn2015} \\
    \hline
  \end{tabular}
\end{table*}

All data was then analysed with a standardized pipeline that includes the \ahf\footnote{\url{http://popia.ft.uam.es/AHF}} \citep{Knollmann2009} halo finder which self-consistently includes both gas and stars in the halo finding process. For each halo, we compute the radius $R_{\rm 200}$, that is the radius $r$ at which the density $M(<r)/(4\pi r^3/3)$ drops below $200\rho_{\rm crit}$\footnote{Similarly, the subscript $500$ used in this paper later are for haloes defined with enclosed overdensities of 500 times the critical density of the Universe.}. Here $\rho_{\rm crit}$ is the critical density of the Universe at the respective redshift. Subhaloes are defined as haloes which lie within the $R_{\rm 200}$ region of a more massive halo, the so-called host halo. As subhaloes are embedded within the density of their respective host halo, their own density profile usually shows a characteristic upturn at a radius $R_t \lsim R_{\rm 200}$, where $R_{\rm 200}$ would be their actual radius if they were found in isolation. We use this ``truncation radius'' $R_t$ as the outer edge of the subhalo and hence subhalo properties (i.e. mass, density profile, velocity dispersion, rotation curve) are calculated using the gravitationally bound particles inside the truncation radius $R_t$. For a host halo which contains the mass of their sub-haloes, we calculate properties using the radius $R_{\rm 200}$. Halo merger trees, that link objects between different redshifts, were constructed using {\sc MergerTree} which forms part of the \ahf\ package. 
We have calculated luminosities in different spectral bands from the stars within the haloes by applying the stellar population synthesis code \textsc{STARDUST} \citep[see][and references therein for more details]{Devriendt99}. This code computes the spectral energy distribution from far-UV to radio, for an instantaneous starburst of a given mass, age and metallicity. The stellar contribution to the total flux is calculated assuming a Kennicutt initial mass function \citep{Kennicutt98}.


The full dataset consists of 324 re-simulated regions, which cover a much larger volume (out to 15 $\Mpc$ in radius) than the central halo's virial radius and hence our sample includes many other objects outside that sphere. These objects are composed of haloes, groups and filaments, which allow us to investigate the preprocessing of the galaxy cluster as well as its large-scale environment. As some of the objects close to the boundary could be contaminated by low resolution particles in the hydrodynamic simulations, we explicitly checked that all the objects included in the comprehensive catalogue do not contain any low resolution particles. In what follows we refer to this dataset, which consists of all the uncontaminated haloes from all the simulations as the `comprehensive' sample (see Appendix \ref{sec:hmf} for details).

\subsection{The semi-analytical models} \label{sec:SAMs}
The aforementioned MDPL2 dark-matter-only simulation has been populated with galaxies by three distinct SAMs, i.e. \galacticus\ \citep{Benson2012}, \sag\ \citep{Cora2018a}, and \sage\ \citep{Croton2016}, and the public release of the resulting catalogues presented in \citet{Knebe2018}. The same 324 regions (using the same radius cut) have also been extracted from the SAMs' halo and galaxy catalogue that covers the entire 1$\Gpc^3$ volume of the parent MDPL2 simulation. This data set constitutes the counterpart sample of the hydrodynamical catalogue, which will be referred as the comprehensive sample as well. This allows for a direct comparison of the same galaxy clusters as modelled by our cosmological simulation codes detailed above. We briefly summarize the salient differences between these SAMs in Table~\ref{table:SAMs}, refering the reader to \cite{Knebe2018} for a more detailed presentation of the three models. Note that \sage\ calculates luminosities in post-processing via the Theoretical Astrophysical Observatory \citep[TAO\footnote{\url{http://tao.asvo.org.au}},][]{Bernyk2016}, which is currently only possible for a sub-volume of the full 1$\Gpc$ box. Therefore, \sage\ will not enter any luminosity-related plots.

\begin{table*}
  \caption{Salient differences between the three SAMs. We only list here whether or not the model has been re-calibrated to the MDPL2 simulation, how it treats orphan galaxies (i.e. galaxies devoid of a dark matter halo), whether it features intra-cluster stars, and how luminosities are available. There are certainly many more differences in the exact implementation of the baryonic physics, but we refer the reader to the model presentation for those details.}
  \label{table:SAMs}
  \begin{tabular}{lllll}
    \hline
    SAM & re-calibration & orphan galaxies & intra-cluster stars & luminosities\\
    \hline
    \galacticus\  & no	& yes, but without positions/velocities & no & yes\\
    \sag\ 	      & yes	& yes, with	full orbit integration & yes & yes\\
    \sage\        & yes	& no	& yes & only for a sub-volume via TAO\\
    \hline
  \end{tabular}
\end{table*}

\section{Cluster bulk properties} \label{sec:bulk}
\begin{figure*}
	\subfloat{\includegraphics[width=0.5\textwidth]{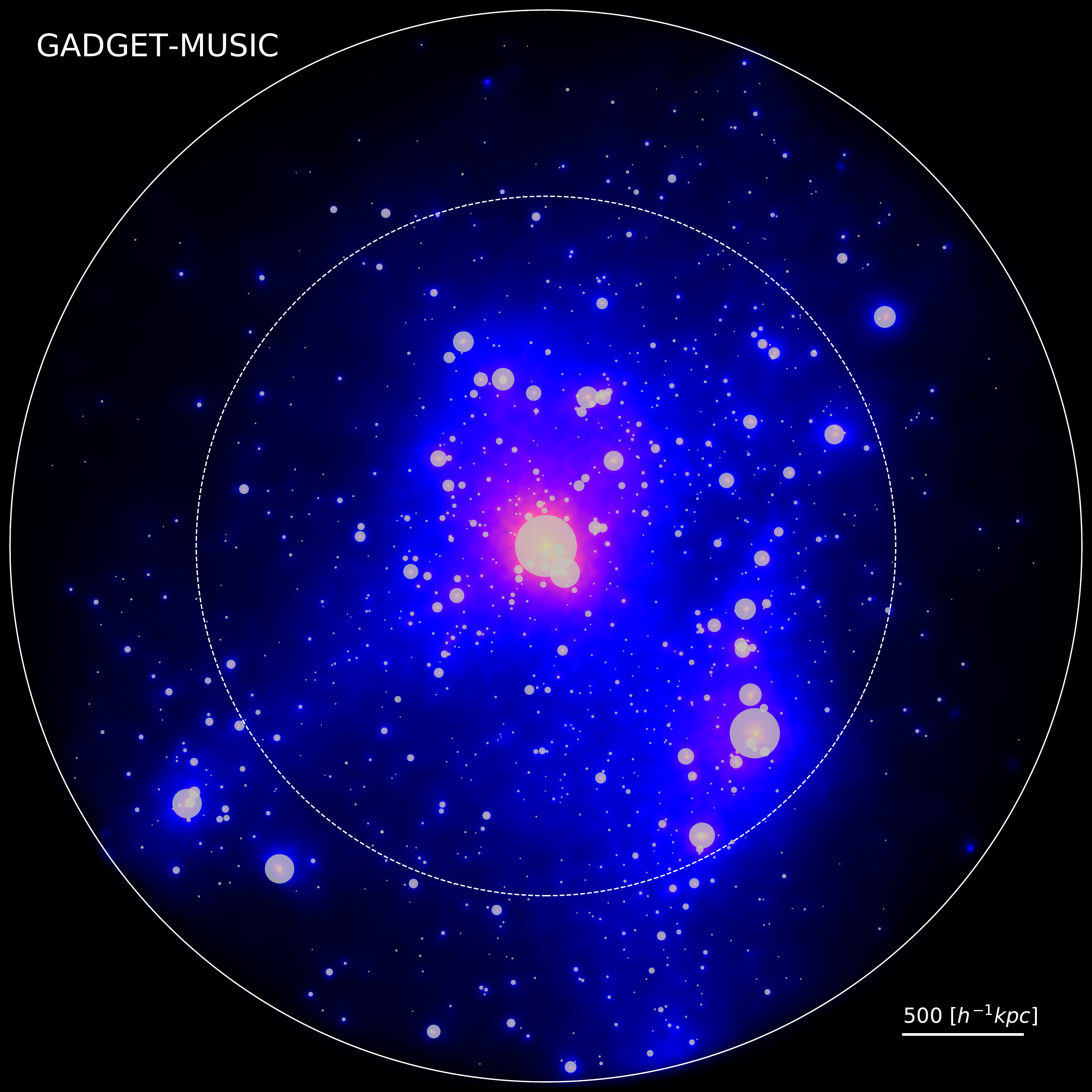}}
    \subfloat{\includegraphics[width=0.5\textwidth]{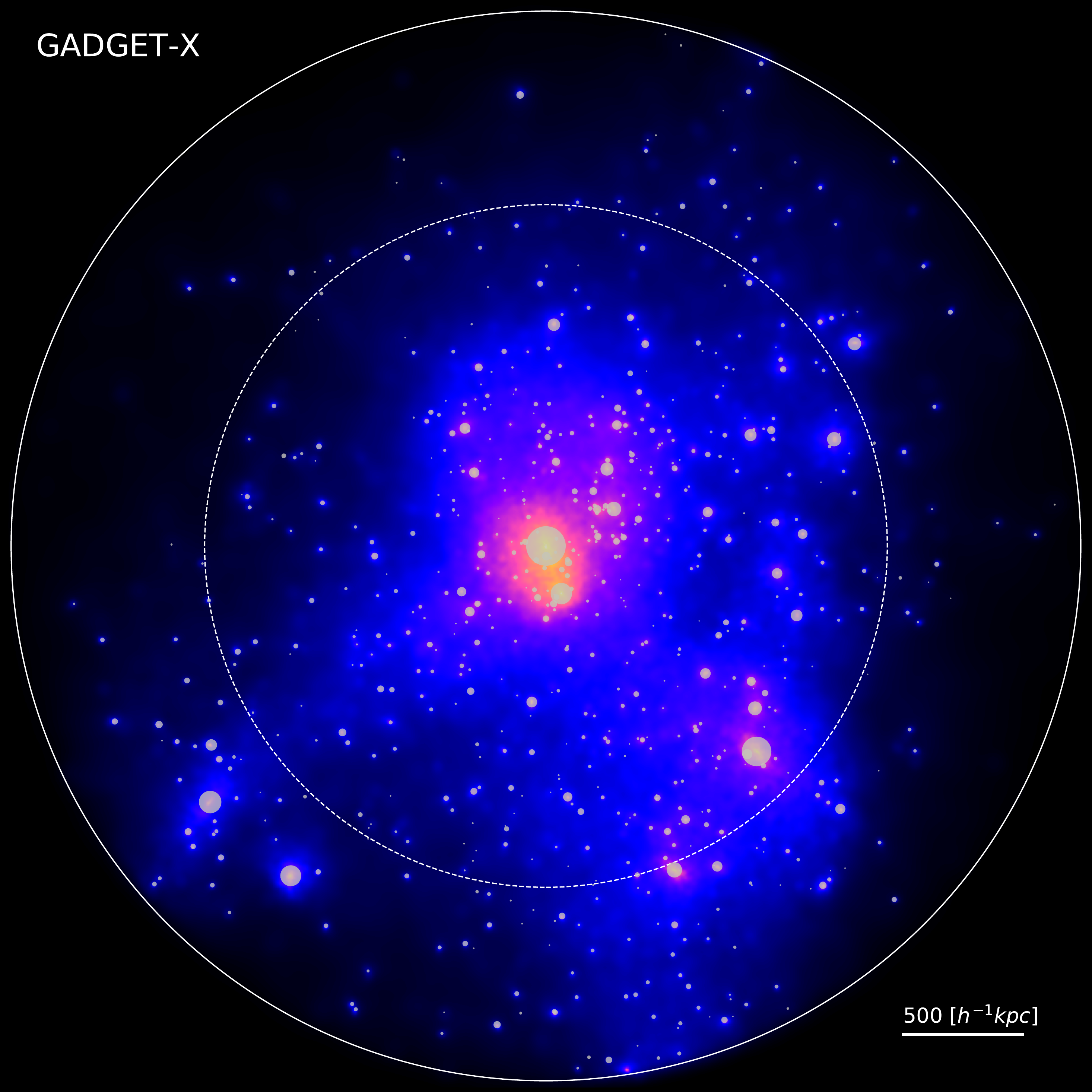}}\\
    \vspace{-1\baselineskip}
    \subfloat{\includegraphics[width=0.5\textwidth]{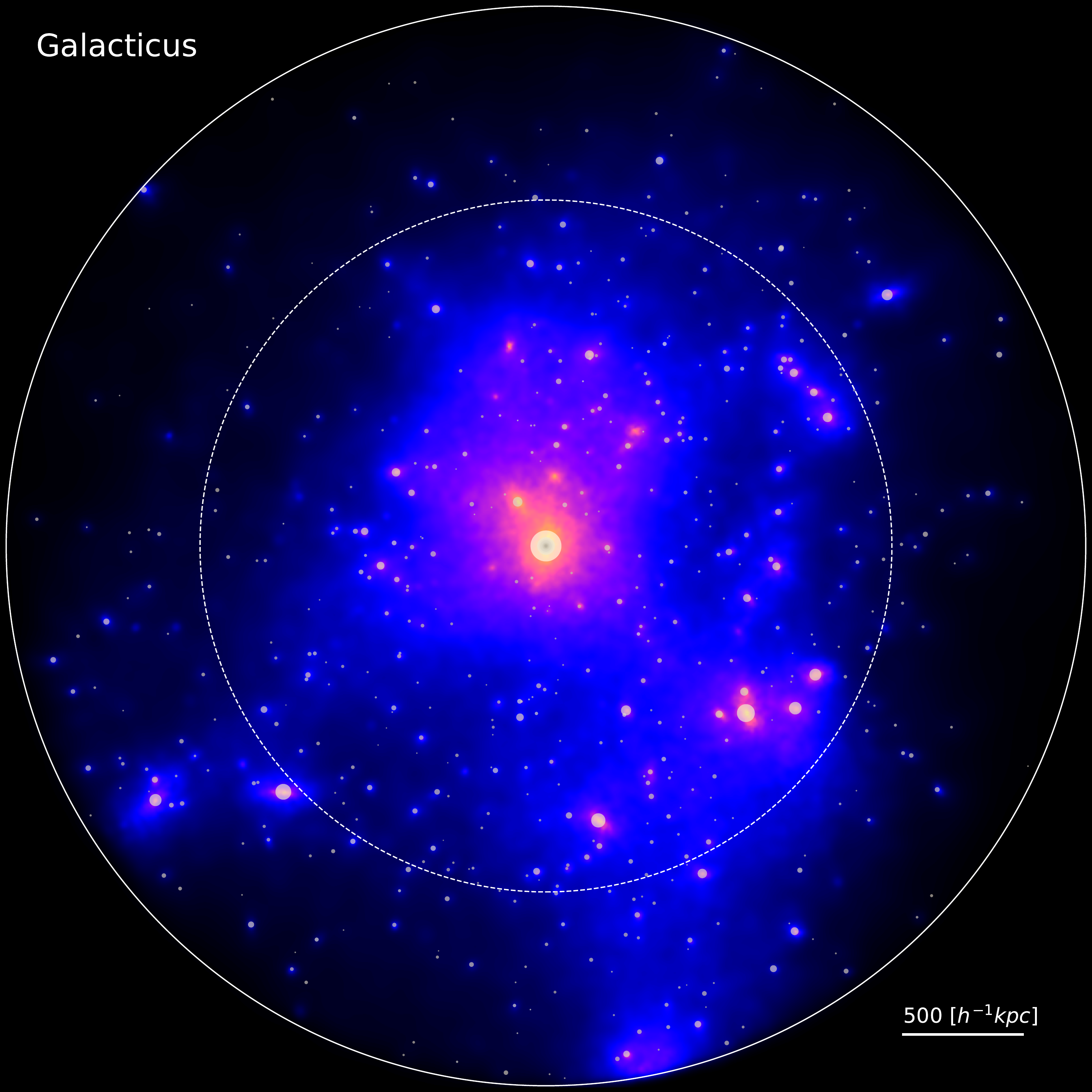}}
    \subfloat{\includegraphics[width=0.5\textwidth]{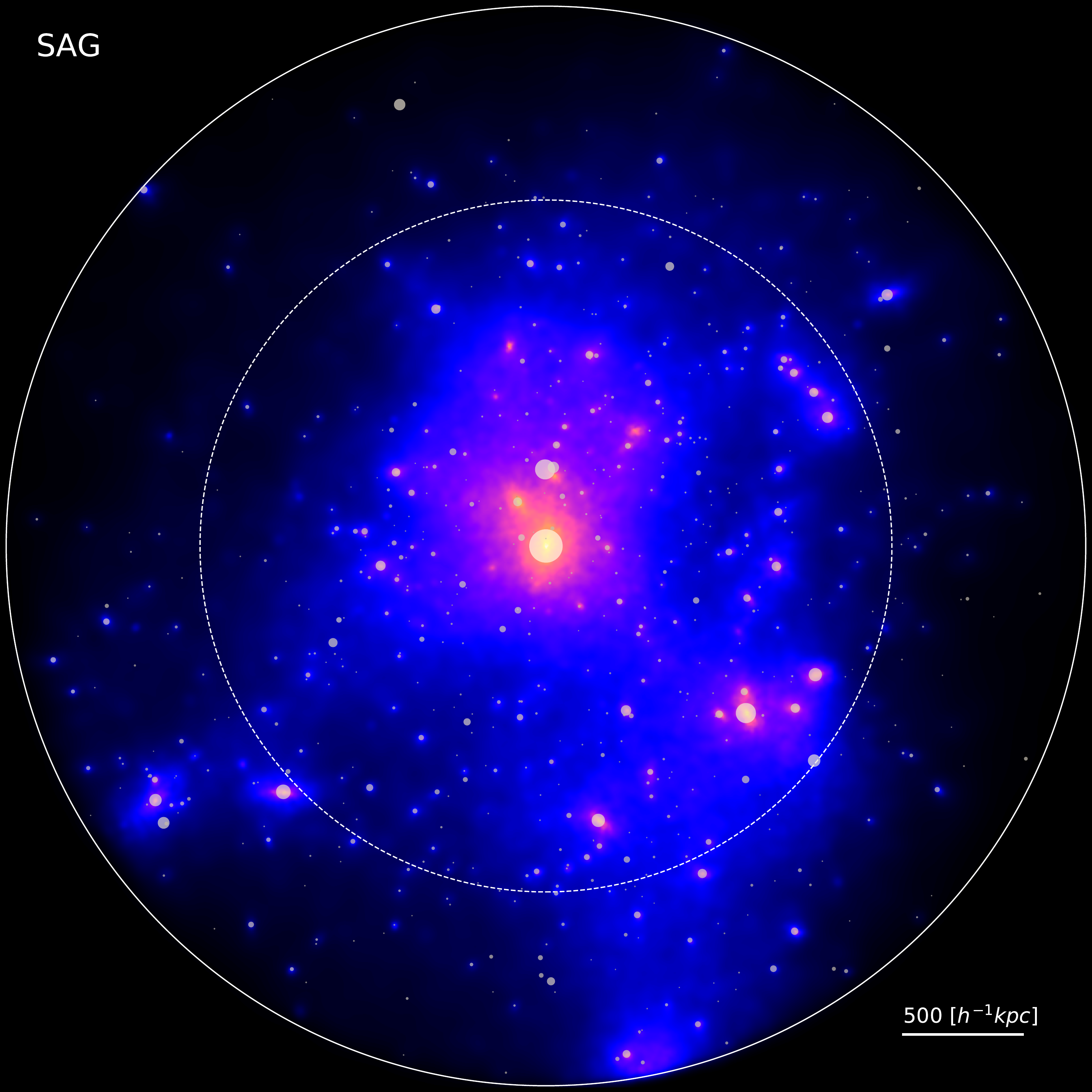}}
	\caption{The distribution of galaxies within $R_{200}$ of the most massive cluster within re-simulation region 1. The upper row shows the results from \gadgetmusic\ (left) and \gadgetx\ (right). The lower row shows the results from the SAMs \galacticus\ (left) and \sag\ (right). The projected dark matter density is shown in the background with a blue-red colour map. Galaxy colour is taken from their SDSS $r$, $g$, and $u$ band magnitude and the symbol size is proportional to stellar mass. The two circles mark the radii $R_{200}$ (outer circle) and $R_{500}$ (inner circle).}
    \label{fig:image}
\end{figure*}

Before quantifying the differences in various cluster properties, we first illustrate in Fig.~\ref{fig:image} the distributions of simulated galaxies and dark matter within a cluster ($r \leq R_{200}$) from one of our re-simulated regions, from both hydrodynamical simulations (upper row) and from SAMs (lower row). Each galaxy is represented by a sphere with size proportional to stellar mass that includes halo stars for the two hydrodynamical simulation, but only uses the stellar mass of the central galaxy for the SAMs. Their colours are based on their SDSS $r$, $g$, and $u$ band luminosities. The background colour map indicates the dark matter density field, which is produced by the {\sc py-sphviewer} code \citep{sphviewer}. The two circles mark the radii $R_{200}$ (outer) and $R_{500}$ (inner). 

It is apparent that the galaxies marked in the different panels are neither exactly in the same position nor do they have the same size for the hydrodynamical simulations. This is not surprising given that the dynamics within the virialised region is non-linear and so small differences in orbit become rapidly amplified. That said, the underlying dark matter density field is visually similar with a large infalling group to the south-east. Both $R_{200}$ and $R_{500}$ are recovered well by the resimulation. The galaxies also differ due to the varying treatment of baryonic processes, as seen in e.g. \cite{Sembolini2016,Sembolini2016a,Elahi2016,Cui2016b,Arthur2017}. Note that the galaxy positions are identical for the two SAMs as they reflect the positions of the dark matter haloes in the underlying dark-matter-only simulation which are the same. The apparent larger sizes for the hydrodynamical galaxies can be related back to the inclusion of halo stars. In agreement with previous studies \citep[for example][]{Ragone-Figueroa2013, Cui2014b, Cui2016b}, the galaxy stellar masses are significantly larger for \gadgetmusic, which does not include a model for AGN feedback.

\subsection{Halo properties}

In this section, we focus on the results from the hydro simulations, noting that the properties of the haloes of the SAM galaxies are identical to the MDPL2 halo properties presented elsewhere \citep[][]{Klypin2016,Knebe2018}


\subsubsection{Baryon effects on halo mass}
\begin{figure*}
	\includegraphics[width=\linewidth]{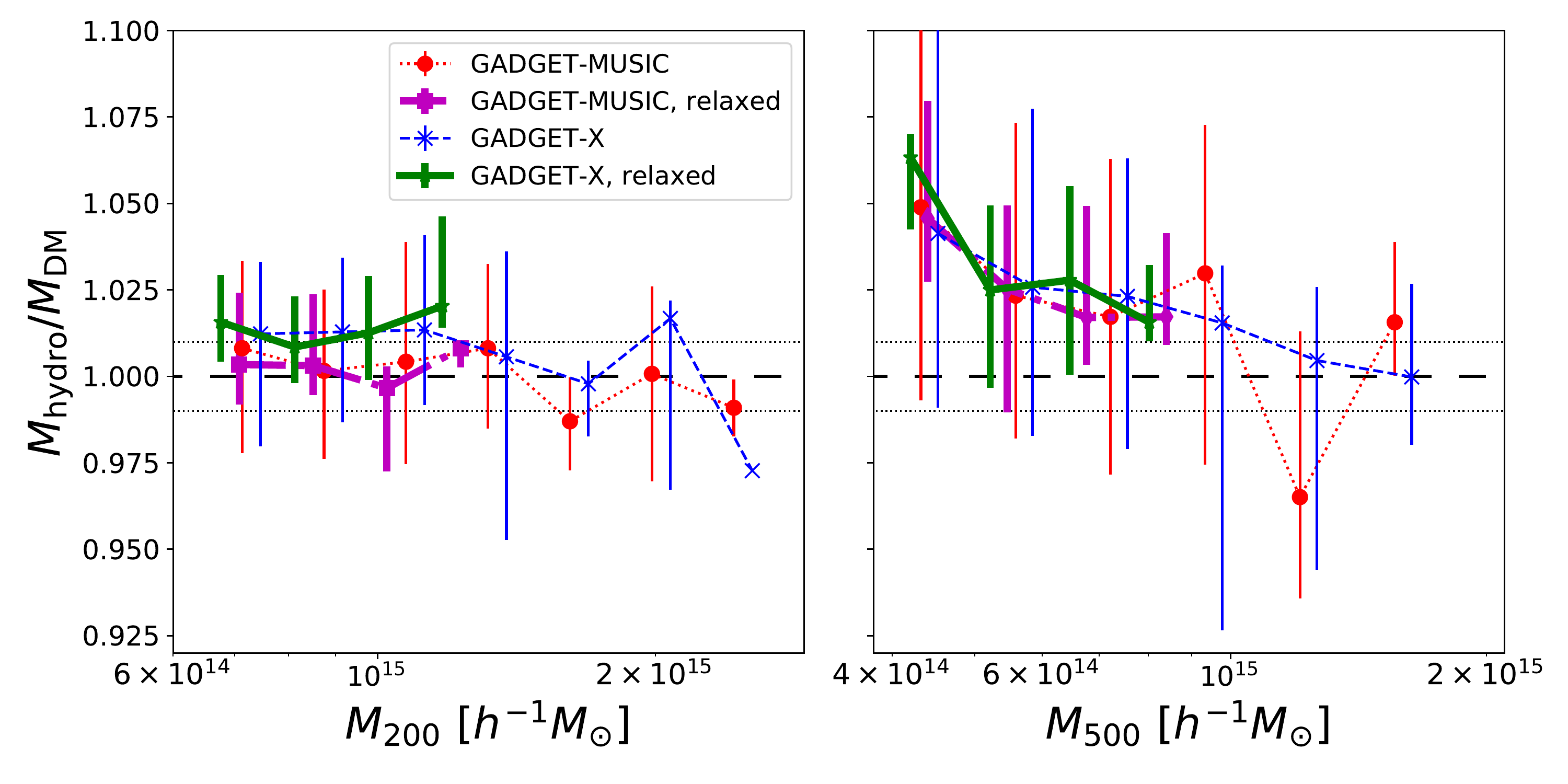}
    \caption{The mass ratio between matched clusters at $z=0$ identified in the hydrodynamical simulations ($M_{\rm hydro}$) and in the corresponding cosmological dark-matter-only run MDPL2 ($M_{\rm DM}$) for $M_{200}$ (left panel) and $M_{500}$ (right panel) as a function of $M_{\rm DM}$. The complete sample used here is in thin lines, while the dynamically relaxed sub-sample is in thick lines. The median value for each mass-bin is shown via the symbols (red dots for \gadgetmusic\ and blue stellar symbols for \gadgetx) with error-bars indicating the $16^{\rm th}$ and $84^{\rm th}$ percentiles. The black horizontal long-dashed and dotted lines indicate equivalent mass and 1 per cent variation respectively.}
    \label{fig:hmd}
\end{figure*}

In order to compare individual clusters between the original MDPL2 simulation and the 324 re-simulated regions the haloes need to be matched. There is generally a direct 1-to-1 alignment between the largest object within the original simulation and the re-simulated region, as illustrated in Fig.~\ref{fig:image}. For the analysis presented here both the original MDPL2 region and the resimulated region have been (re)processed using \ahf. This ensures exact consistency between the halo finder definitions, i.e. it avoids effects introduced by using results from different halo finders \citep{Knebe2011,Knebe2013}. Further, \ahf\ can extract haloes self-consistently from simulations including gas and stars as well as dark matter. We use the halo centre position as the primary criteria for matching the clusters and select the one with the nearest mass when there are multiple matches. As previously mentioned the exact halo positions will have moved slightly from those in the original dark-matter only simulation but these changes are generally small (at the level of a few percent of the virial radius in most cases, \citealt{Cui2016b}). Occasionally the differences are larger, typically due to the presence of an ongoing merger. It has been shown that halo finders struggle to uniquely track the main halo through a merger and rather treat the two participating objects as a host-subhalo system \citep{Behroozi2015}. Furthermore, the cluster centre can flip between different density peaks (subhaloes) due to baryonic processes \citep{Cui2016a}. That said, in our worst-case scenario, we have two matched haloes with $\sim 40$ per cent mass difference caused by a massive merging subhalo. In general cases, these different kinds of mismatching only happen for the dynamically un-relaxed clusters, not for the relaxed ones.

Accurate estimates of cluster masses are very important for constraining cosmological parameters and cosmological models \citep[for example,][]{Bocquet2016, Sartoris2016}. Therefore, we present here a quantitative comparison of the halo masses as found in the hydrodynamical simulations with their respective counterparts from the dark-matter-only MDPL2 simulation \citep[see][for a review of the baryon effect]{CuiBC2017}. \Fig{fig:hmd} shows the mass ratio of clusters in \gadgetmusic\ (red circle and lines) and \gadgetx\ (blue star and lines) to their MDPL2 counterparts; $M_{200}$ is shown in the left-hand side panel and $M_{500}$ in the right-hand side panel.\footnote{The $M_{500}$ sample was constructed by using \ahf\ to find the largest halo contained within each of the 324 clusters of the mass-complete sample (and matching these as before).} In order to reduce any issues due to mismatching, we select a sample of dynamically relaxed clusters (see below for details) from the complete sample and repeat the comparison. The mass ratio for $M_{200}$ from both hydrodynamical simulations is very close to unity (with the median difference lying basically within 1 per cent), with a scatter less than $\sim 5$ per cent ($\sim 2.5$ per cent for the relaxed sub-sample). At the low mass end, \gadgetx\ (for both samples) tends to have about 1 per cent higher mass than its MDPL2 counterpart. However, the $M_{500}$ mass in both sets of hydrodynamical simulations tends to be several (up to 6) per cent higher than its dark-matter only counterpart below $\sim 9\times10^{14}\hMsun$. Above this halo mass the ratio drops to around 1 again. It is worth noting that for $M_{500}$ there is a larger scatter of $\sim 8$ per cent for the complete sample and $\sim 4$ per cent for the relaxed sub-sample. We ascribe this larger mass change for $M_{500}$ to baryonic processes which have a larger effect closer to the cluster's centre and for the less massive haloes. The two simulation codes show similar results for $M \gsim 10^{15} \hMsun$ at both overdensities, which means that the baryon physics has little influence on both $M_{500}$ and $M_{200}$ at this cluster mass range. For the $M_{200}$ mass changes, this is in agreement with previous similar comparisons \citep[e.g.][]{Cui2012a,Cui2014a,Cui2016b}. For $M_{500}$, \cite{Cui2014a} reported a slight mass decrease when AGN feedback is included and a slight mass increase without AGN feedback. At this halo mass range, $M_{500} > 10^{14.5}\hMsun$, the difference between \gadgetx\ and \cite{Cui2014a} could be caused by either a sample effect (\citealt{Cui2014a} studied very few clusters) or due to the details of the baryonic model implemented in the simulation. We will explore this in detail in a follow-up paper (Cui et al. in prep.) which will also focus on cluster mass estimates based upon different observational methods applied to our simulation data.

\subsubsection{Dynamical Relaxation}
\begin{figure*}
	\includegraphics[width=\linewidth]{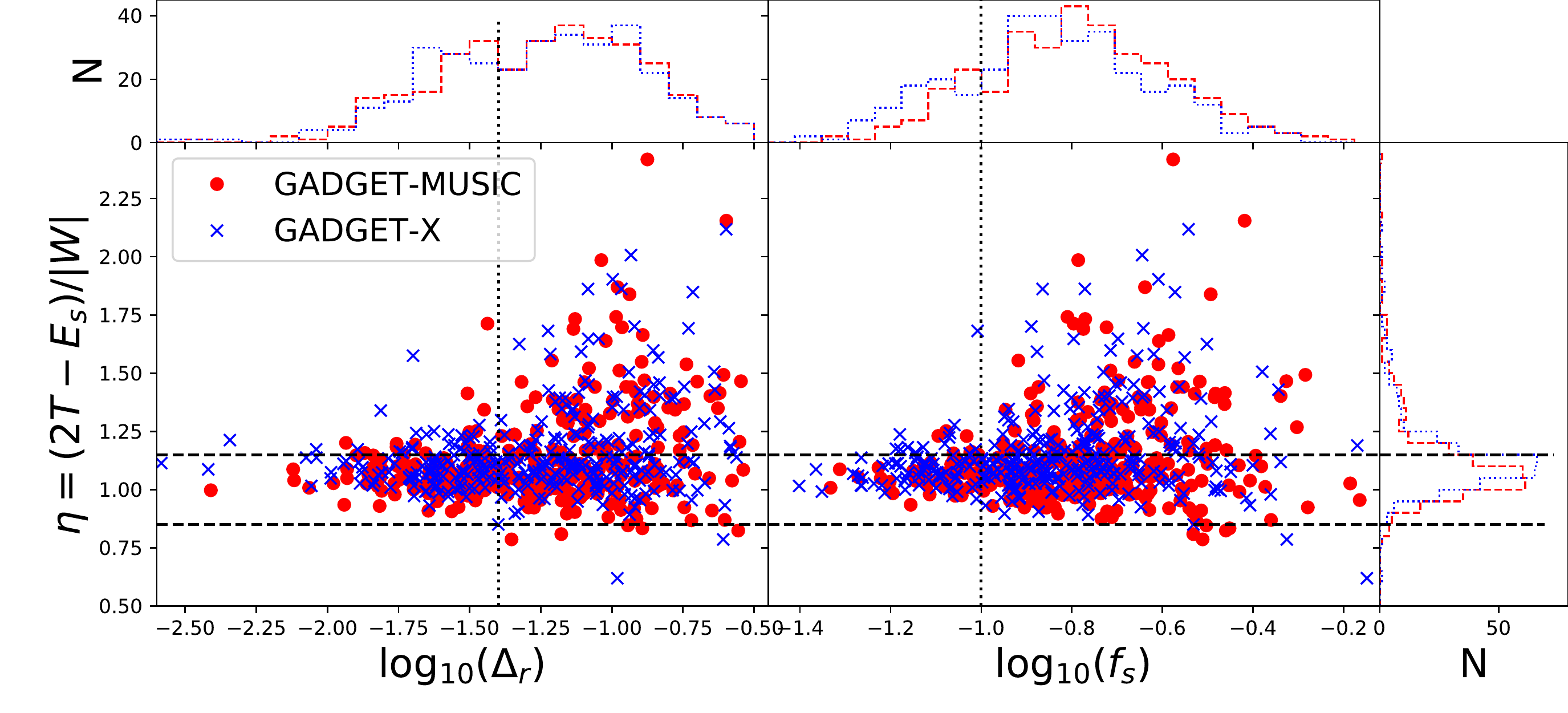}
    \caption{For the mass-complete sample, the left hand panel shows the relation between the virial ratio ($\eta$) and the centre-of-mass offset ($\Delta_{\rm r}$). The right-hand side panel shows the relation between $\eta$ and the subhalo mass fraction ($f_{\rm s}$). The top and right-hand sub-panels show their corresponding histograms. Red filled circles (red dashed line for the histogram) show the clusters from the \gadgetmusic\ run, while the blue crosses (blue dotted line for the histogram) show the \gadgetx\ results. The two horizontal dashed lines show the selection limits for the $\eta$ parameter, while the vertical dotted lines show the selection limits for $\Delta_{\rm r}$ and $f_{\rm s}$ (see text).}
    \label{fig:ds}
\end{figure*}

\begin{table}
	\centering
	\caption{The fraction of relaxed clusters. The first column shows the mass range. The second to fourth columns show the relaxation fractions from; all three methods combined, $\Delta_r$ plus $f_{\rm s}$, and only $f_{\rm s}$ criterion, respectively. Each cell shows two values, of which the first one is the relaxation fraction for \gadgetmusic\ and the second value is for \gadgetx. Clusters with $M_{200} < 6.42 \times 10^{14} \ \hMsun$ (mass bins above the dashed line) are taken from the comprehensive sample.}
	\label{tab:relaxation}
	\begin{tabular}{lccc} 
		\hline
		$M_{200}$ [$10^{14} \hMsun$] & $\eta$, $\Delta_{\rm r}$ \& $f_{\rm s}$ & $\Delta_{\rm r}$ \& $f_{\rm s}$ & $f_{\rm s}$\\
		\hline
		$0.10 - 0.50$		& 0.44 / 0.36 & 0.56 / 0.48 & 0.70 / 0.65\\
		$0.50 - 1.00$		& 0.36 / 0.34 & 0.45 / 0.46 & 0.56 / 0.57\\
		$1.00 - 6.42$ 		& 0.27 / 0.29 & 0.30 / 0.35 & 0.43 / 0.48\\ \hdashline
        $> 6.42 $ 			& 0.15 / 0.17 & 0.16 / 0.21 & 0.17 / 0.23\\
		\hline
	\end{tabular}
\end{table}

To determine the dynamical state of the cluster sample we study three indicators, following \cite{Cui2017}, specifically:
\begin{itemize}
\item the virial ratio $\eta=(2T - E_{\rm s})/|W|$, where $T$ is the total kinetic energy, $E_{\rm s}$ is the energy from surface pressure and $W$ is the total potential energy,
\item the centre-of-mass offset $\Delta_{\rm r}=|R_{\rm cm} - R_{\rm c}|/R_{200}$, where $R_{\rm cm}$ is the centre-of-mass within a cluster radius of $R_{200}$, $R_{\rm c}$ is the centre of the cluster corresponding to the maximum density peak of the halo. Using the position of the minimum of the gravitational potential would give a similar result as investigated by \cite{Cui2016a}.
\item the fraction of mass in subhaloes $f_{\rm s}=\sum M_{\rm sub}/M_{200}$ where $M_{\rm sub}$ is the mass of each subhalo.
\end{itemize}
We adopt the following criteria to select dynamically relaxed clusters: $0.85 < \eta < 1.15$, $\Delta_{\rm r} < 0.04$ and $f_{\rm s} < 0.1$, which need to be satisfied at the same time \citep[see, for instance,][]{Neto2007,Knebe2008,Power2012}. Note that we use here a slightly larger limit for $f_{\rm s}$ than in \cite{Cui2017}. This is because (1) $R_{200}$ is used instead of the virial radius\footnote{Note that for the given cosmology $R_{200}<R_{\rm vir}$ and hence the $M_{200}$ masses of the host haloes considered here will be about 25 per cent smaller than $M_{\rm vir}$.}, and (2) this threshold for $f_{\rm s}$ gives a relaxation fraction ($\sim 20$ per cent for both hydrodynamical simulations) comparable to observations \citep[for example,][]{Mantz2015,Biffi2016}.

In Fig.~\ref{fig:ds}, we show the relations between these three parameters for the mass-complete sample: $\Delta_{\rm r}$ versus $\eta$ in the left-hand panel and $f_{\rm s}$ versus $\eta$ in the right-hand panel. The two hydro-runs show a similar distribution of relaxed clusters (shown for convenience at the top and to the right of the figure panels), in agreement with \cite{Cui2017}. The histogram peak of the $\eta$ parameter from \gadgetx\ has a slightly higher value than the peak from \gadgetmusic. This could be due to the AGN feedback, which releases additional energy into the kinetic component. 

A quantitative analysis of the relaxation fraction within our comprehensive halo catalogue, for different mass bins and with different combinations of relaxation parameters is given in Table~\ref{tab:relaxation}. The fraction of relaxed clusters shows a clear decreasing trend as halo mass increases. This is simply because the more massive the object is, the less likely it is to have reached dynamical relaxation by redshift $z=0$. This can be traced back to the relation between formation time and halo mass \citep[see Fig.~2 in][for instance]{Power2012}. There is very little change in relaxation fraction for the complete sample when different criteria are applied. There is a noticeable difference in the relaxed cluster fraction for the smallest mass bin, with the fraction for \gadgetx\ being significantly lower ($\sim 8$ per cent) than that for \gadgetmusic\ when all three criteria are applied. This is due to the AGN feedback in \gadgetx\ efficiently ceasing star formation in small objects and creating gas turbulence. The relaxation fractions for the mass-complete sample from both \gadgetmusic\ and \gadgetx\ show an obvious decrease. On the contrary to the smallest mass bin, the relaxation fraction from \gadgetmusic\ seems lower than from \gadgetx. This overturn is simply because the mass fraction of substructures in \gadgetmusic\ is higher than \gadgetx, which dominates the relaxation fraction. In an upcoming paper we will provide a more detailed investigation of the evolution of the cluster dynamical state and the impact of input physics on various observational classification methods (Old et al. in prep.).

\subsubsection{Concentration-Mass ($c-M$) relation}
\begin{figure*}
	\includegraphics[width=\linewidth]{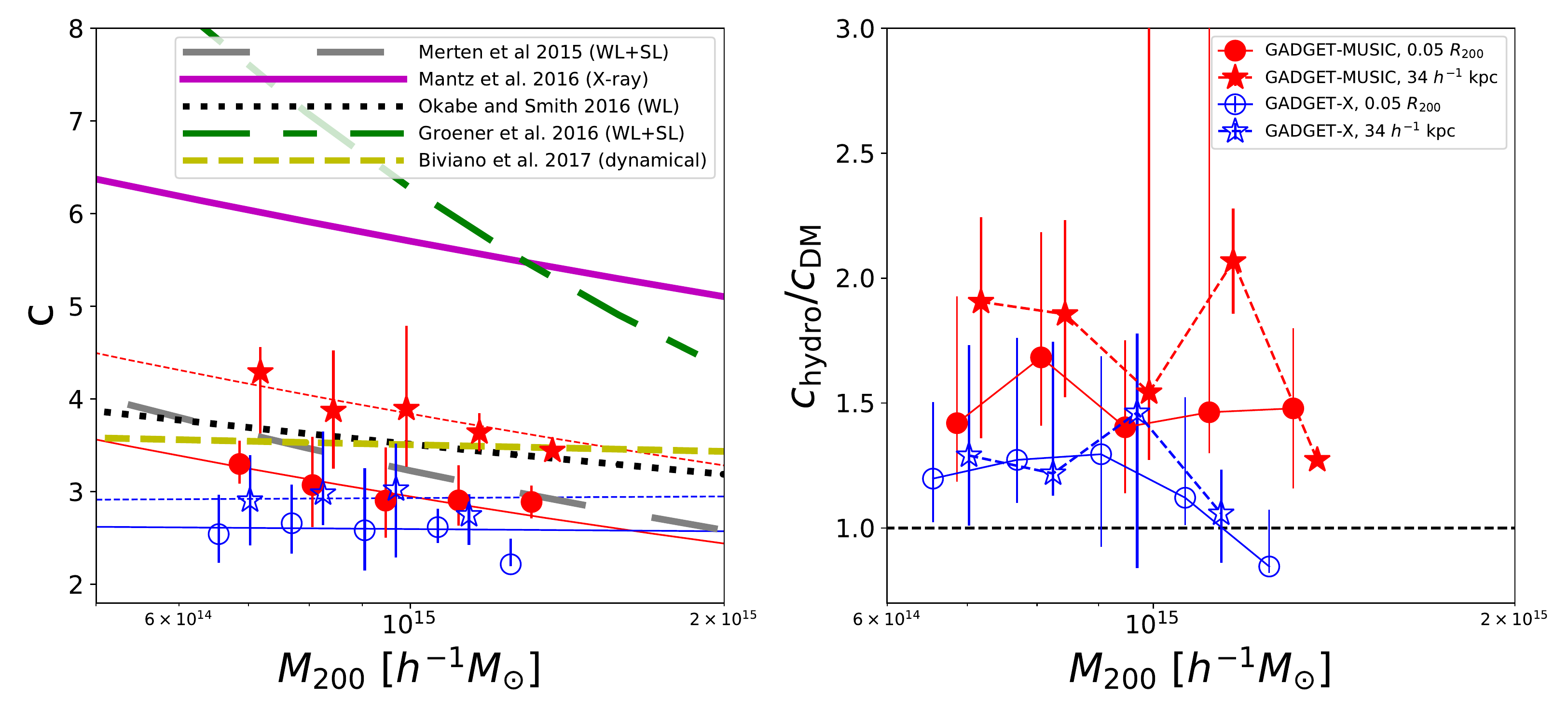}
    \caption{Left panel: The concentration--halo mass relation for the relaxed galaxy clusters from the two hydrodynamical simulation runs compared with various observational results. As indicated in the legend, thick lines with different styles show the best fit results from recent observational data obtained with different methods \citep{Merten2015,Mantz2016,Okabe2016,Groener2016,Biviano2017}. Symbols show the median values with the $16^{\rm th}$ - $84^{\rm th}$ percentile error-bars from the hydro simulations: circles and stars (red filled symbols for \gadgetmusic\ and blue open symbols for \gadgetx) for the concentration derived by fitting the density profile up to two inner radii ($34 \Kpc$ and $0.05\ R_{200}$, see text for details). The red (blue), thin solid and dashed lines are the best fit result to the concentration mass relation of \gadgetmusic\ (\gadgetx) clusters. In the right panel of this figure, we represent the ratio of the concentration between the hydrodynamical simulation clusters and their match in the original MDPL2 dark-matter-only simulation. Again, the symbols show the median values with the $16^{\rm th}$ - $84^{\rm th}$ percentile error-bars.}
    \label{fig:cm}
\end{figure*}

\begin{table}
	\centering
	\caption{The fitting parameters for the concentration-mass relation with fitting function: $\log_{10} c_{200} = \alpha - \beta\log_{10} M_{200} / M_{\odot}$. The first row shows the results with the inner radius set to $0.05\ R_{200}$, while the second row is for a $34 \Kpc$ inner radius. Each cell shows two values, of which the first one is for the fitting parameter $\alpha$ and the second value is $\beta$.}
	\label{tab:cmp}
	\begin{tabular}{lcc}
		\hline
		Inner radius & \gadgetmusic\ & \gadgetx\ \\
        		  & $\alpha$ / $\beta$ & $\alpha$ / $\beta$\\
		\hline
		$0.05\ R_{200}$ & 4.60 / 0.27 & 0.62 / 0.013\\
		$34 \Kpc$ 	& 4.02 / 0.23 & 0.34 / -0.01\\
		\hline
	\end{tabular}
\end{table}

Knowledge of the halo concentration, $c$, and mass, $M$, would specify the full evolution of a halo in the spherical collapse model \citep{Bullock2001}. The relation $c-M$ between these two fundamental properties, alongside its standard deviation, are related to the variance in the assembly histories of dark matter haloes \citep[e.g.][]{Zhao2003a, Zhao2003b}. Furthermore, the normalization and evolution of this relation also depend on the cosmological model \citep[e.g.][]{Dolag2004,Carlesi2012}. However, there exists some tension between the observationally estimated relation and the theoretical prediction. This could result from not comparing like-with-like when contrasting baryonic simulations and observational results with carefully imposed selection criteria \citep[see][for example]{Rasia2013,Biviano2017}. Here, we only use our relaxed galaxy clusters from the mass-complete sample to investigate and compare this relation with the observational results.

The halo density profiles can be analytically described by an NFW profile \citep{Navarro1997}, 
\begin{equation}
\frac{\rho(r)}{\rho_{\rm crit}} = \frac{\delta_{\rm c}}{(r/r_{\rm s})(1+r/r_{\rm s})^2},
\end{equation}
which is characterized by the two parameters, $r_{\rm s}$ and $\delta_{\rm c}$. The concentration $c_{200}$ is then given by $R_{200}/r_{\rm s}$. We fit our simulated cluster density profiles, defined by equally spaced log-bins, to this functional form with both parameters free, but exclude the very central region in this process. Due to the presence of the BCG, the mass profile in the centre is much steeper than the total mass profile \citep{Schaller2015b}. As the edge of the BCG is not clearly defined, we adopt two different inner `exclusion' radii during the fitting: $0.05\ R_{200}$, as suggested by for example \cite{Schaller2015b,Cui2016b} and $\sim 34 \Kpc$ following \cite{Biviano2017}. We have verified that the NFW profile provides a good fit regardless of the adopted inner radii ($34 \Kpc$ or $0.05\ R_{200}$). In both cases the difference between the fit and the original density profile is within $20$ per cent at all radii.

In the left panel of \Fig{fig:cm} we show the $c-M$ relation for our relaxed galaxy clusters and compare the relation with observational results coming from both X-ray and optical data obtained with different techniques (please refer to the figure caption and legend, respectively). For each of the two hydrodynamical simulation codes, we show results stemming from either truncation approach: circles for using the range $[0.05\ R_{200} - R_{200}]$ and stars for a fixed inner radius of $34 \Kpc$. We fit our $c-M$ relation using the following analytical function:
\begin{equation}
\log_{10} c_{200} = \alpha - \beta\log_{10} (M_{200}/M_{\odot})
\end{equation}
The fitting parameters $\alpha$ and $\beta$ are listed in table \ref{tab:cmp}.

It is evident that the $c-M$ relation from our hydro-simulated clusters is closer to the observational results from \cite{Merten2015,Okabe2016,Biviano2017} than those from \cite{Mantz2016,Groener2016}. The $c-M$ relation from the \gadgetmusic\ run is slightly higher than from the \gadgetx\ run and it is in better agreement with observational results which have lower concentrations. It is obvious that the concentrations with a $34 \Kpc$ inner cut-off are systematically higher than the ones with a $0.05\ R_{200}$ cut-off \citep[see also][for similar results with different inner radii]{Rasia2013}. Our fitted $c-M$ relation from the \gadgetx\ clusters is much flatter than \cite{Schaller2015a}, simply because their fit covers a much larger mass range, which is dominated by the lower mass objects. Furthermore, \gadgetx\ shows an increasing slope with $\beta = -0.01$ when a fixed inner radius of $34 \Kpc$ is taken. This can be understood because $34 \Kpc$ corresponds to a smaller fraction of $R_{200}$ for a massive cluster than for a less massive halo. Therefore it is not surprising to see a relatively high concentration for the most massive haloes when a fixed physical cut-off radius is applied.

In the right panel of \Fig{fig:cm}, we investigate the baryon effects on the $c-M$ relation by showing the relative change in concentration from dark-matter-only simulated clusters to their equivalent in the two hydro-runs. The change on $c-M$ relation due to baryons varies from $\sim 25$ per cent (for both radii) for \gadgetx\ to about 1.5 - 2 times ($0.05\ R_{200}$ - $34 \Kpc$) for \gadgetmusic. However, this ratio is much lower for the highest mass bin for \gadgetx\ with both inner radii (also for \gadgetmusic\ with the inner radius of $34 \Kpc$). The influence of baryons on the concentration is a little higher than in \cite{Rasia2013}, which may be the result of both the different radius range used for profile fitting and differences in the baryonic model employed.

\subsection{Baryon fractions}

\begin{figure*}
	\includegraphics[width=\linewidth]{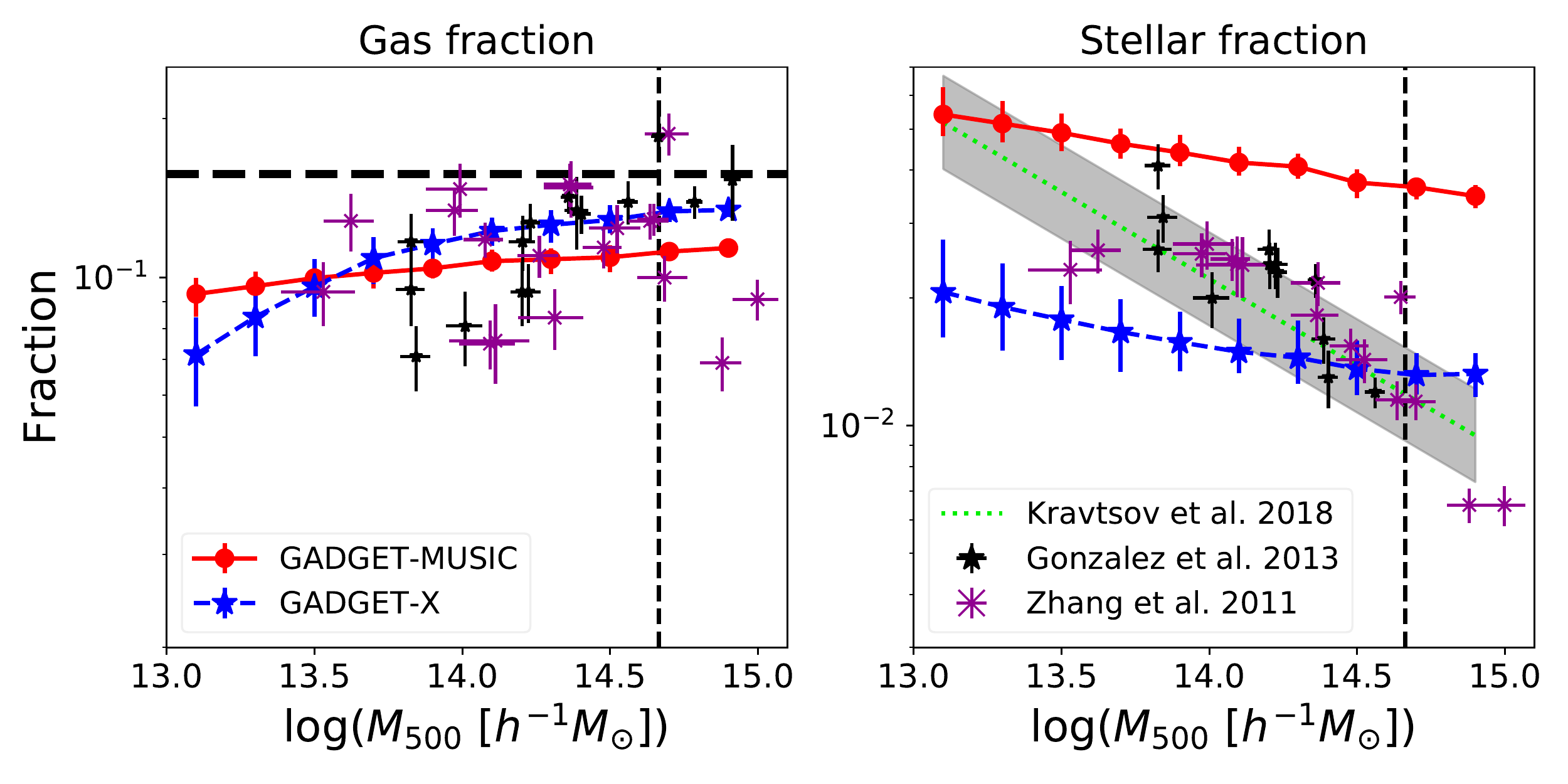}
    \caption{The baryonic fractions from the two hydrodynamical simulations within {\it $R_{500}$}. Gas fractions are shown on the left-hand side panels, while stellar fractions are shown on the right-hand side panels. As shown in the legend on the top-left panel, hydrodynamical simulations are shown with red filled symbols (median value) with error-bars ($16^{\rm th} - 84^{\rm th}$ percentile) for \gadgetmusic\ and blue stars with error-bars for \gadgetx. Observational data points from \citet{Gonzalez2013} and \citet{Zhang2011} are shown as black stars and magenta cross symbols respectively, while the lime dotted line shows the fitting result from \citet{Kravtsov2018} with the grey shaded scatter. The thick black horizon dashed lines on the left-hand side panels indicate the cosmic baryon fraction ($\Omega_{\rm b}/\Omega_{\rm m}$). The vertical dashed lines in the upper row shows the mass limit for the complete sample.}
    \label{fig:bf}
\end{figure*}

The formation of a galaxy cluster depends not only on gravity acting on cosmic scales but also on sub-resolution phenomena such as star formation and various feedback mechanisms returning energy back to the intra-cluster gas. It is a process that involves interplay between dark and baryonic matter. One of the most important quantities to quantify the relation between dark matter and baryons is the baryonic mass fraction. It has therefore been intensively studied: on the theoretical side, mostly by means of hydrodynamical simulations \citep[e.g.][]{MUSICI,Planelles2013,Wu2015,Barnes2017b}; on the observational side via multi-wavelength observations \citep[e.g.][]{Lagana2013,Eckert2016,Chiu2017}. 


In Fig.~\ref{fig:bf}, we show the gas and stellar mass fractions for the comprehensive sample from hydrodynamical simulations within $R_{500}$.
The gas fraction for \gadgetx\ is larger than for \gadgetmusic\ at the massive end, and drops more quickly towards lower mass haloes. The gas fraction from \gadgetx\ shows a better agreement with the data of \citet{Gonzalez2013} at the massive end; both simulations are in line with the results from \citet{Zhang2011} due to its large scatter. The offset between the two hydro-runs is much larger (a factor of 2 - 3) for the stellar fraction. Again, \gadgetx\ shows a better agreement with the observational data points at the massive end. However, it has a flatter slope than the observational results, which is close to the \gadgetmusic\ result at $M_{500} \lesssim 10^{13.5} \hMsun$. This is possibly caused by the strong AGN feedback in \gadgetx. Essentially both hydrodynamic models have a stellar fraction versus mass slope that is inconsistent with the observational data.


Previous comparisons of the stellar and gas mass fractions from full-physics hydrodynamical simulations with observations have shown that models without AGN feedback consistently have too low a gas fraction and too high a stellar fraction due to the over-cooling problem \citep[for example][]{Planelles2013}. This is also seen in \Fig{fig:bf} comparing the \gadgetmusic\ and the \gadgetx\ runs. Although \gadgetx\ tends to have a better agreement with the observational results, the AGN feedback implementation featured by this code is still not perfect: the most massive clusters at $M_{500} \gtrsim 10^{15} \hMsun$ still have a stellar fraction that is a little too high; while intermediate and low mass haloes ($M_{500} \lesssim 10^{14} \hMsun$) have stellar fractions that are too low. Nevertheless, we note that the stellar mass fraction estimated from observations is not without issues: there is relative uncertainty about the contribution of the intra-cluster light \citep[for example][]{Zibetti2005,Gonzalez2007,Puchwein2010,Cui2014b}, which is included in \cite{Gonzalez2013} and \cite{Kravtsov2018}, but not in \cite{Zhang2011}; another problem is the influence of the different initial mass functions adopted in observations to derive stellar mass from luminosities \citep[see e.g.][ for detailed discussions]{Chiu2017}.

The difference in the stellar mass fractions shows the importance of the detailed prescription for baryon processes. Therefore, we are working on a follow-up paper (Rasia et al., in prep.) to investigate in detail the connection between the encapsulated physics and the resultant baryonic fractions, examining the difference between relaxed and un-relaxed clusters, between cool core and non-cool core clusters, as well as the redshift evolution of these fractions.

\section{Stellar and Gas relations of Clusters} \label{sec:sr}

Scaling relations between the total cluster mass and observational quantities are derived in several multi-wavelength studies. Commonly used observational probes include stellar luminosity, X-ray temperature or the Comptonization parameter \citep[e.g.][]{Reiprich2002,Lin2004a,Andersson2011}, which normally show a self-similar relation to cluster mass. They are very powerful tools to derive total cluster masses from different observations. Before this can happen, they need to be accurately calibrated and their dispersions properly estimated. It is worth noting that the scaling relations derived from observations could be biased by sample selection which should have no influence on our mass-complete sample. In this section, we investigate the scaling relations found in our hydrodynamical simulations, and compare them with those from SAMs and observations.

\subsection{Stellar relations}
\subsubsection{Stellar-to-halo mass relation}
\begin{figure}
	\includegraphics[width=\linewidth]{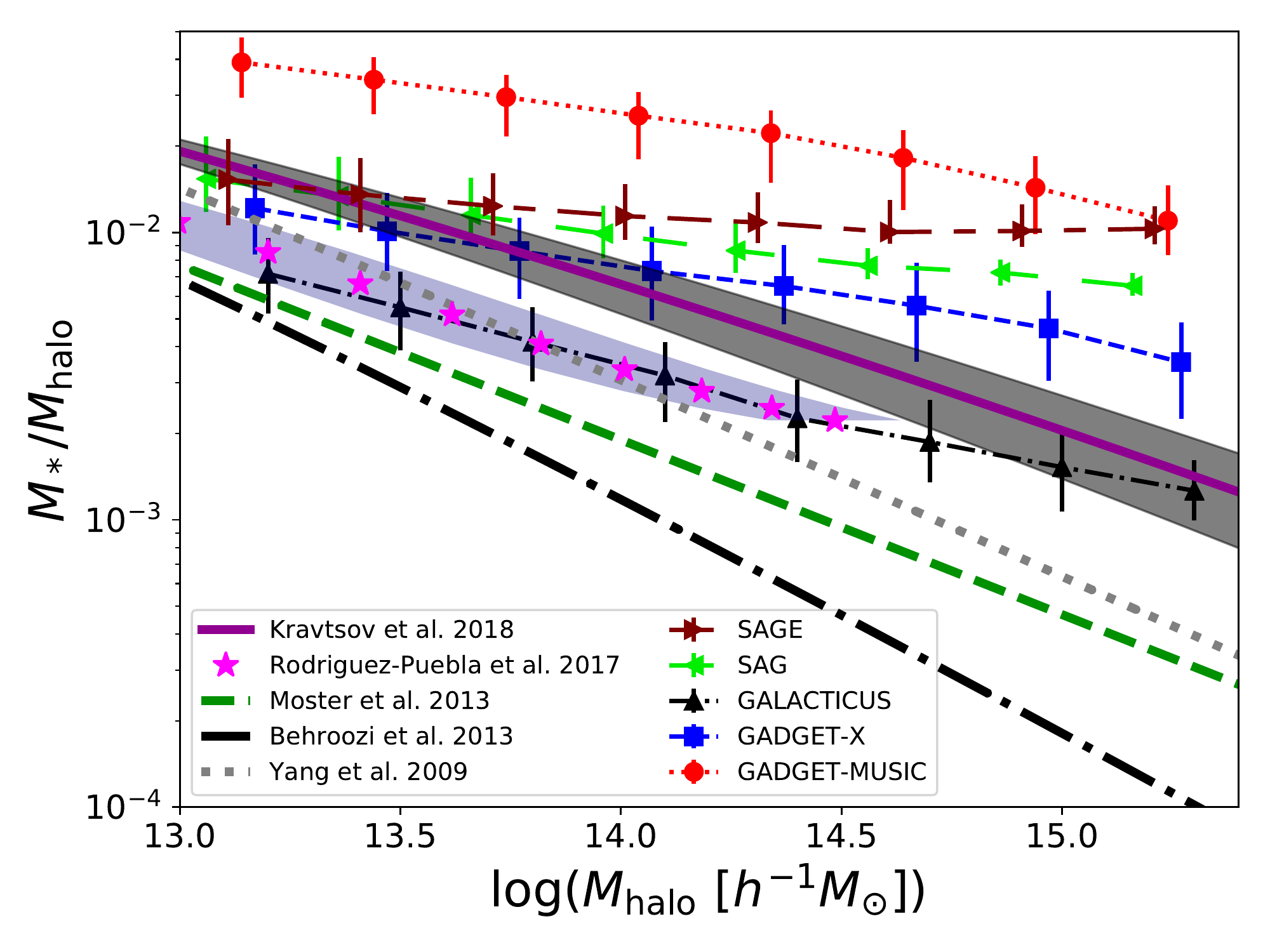}
    \caption{The stellar-to-halo mass relation for central galaxies in the complete sample. As indicated in the legend, observational results are shown as thick lines (\citet{Yang2009}, grey dotted line, \citet{Behroozi2013}, dot-dashed black line and \citet{Moster2013}, green dashed line) with the latest results from \citet{Rodriguez-Puebla2017} shown as magenta stars with the light shaded area and \citet{Kravtsov2018} as a solid purple line with the dark shaded region. Our hydrodynamical simulation and SAM results are shown in different symbols (median value) with error-bars ($16^{\rm th} - 84^{\rm th}$ percentile): \gadgetmusic\ with red solid circles and dotted line; \gadgetx\ with blue solid squares and dashed line; \galacticus\ with black filled triangles and dash-dotted line, \sag\ with lime triangles and long dashed line and \sage\ with maroon triangles and long-short dashed line.}
    \label{fig:shr}
\end{figure}

How galaxy properties relate to their host dark matter halo is an open question in astronomy. Therefore, a substantial effort has focused on establishing robust determinations of the galaxy-halo connection, commonly reported in the form of the stellar-to-halo mass relation, SHMR \citep[][and references therein]{Guo2010,Yang2012,Moster2013,Behroozi2013}. In Fig.~\ref{fig:shr}, we compare our SHMR with results from the literature. It is worth noting here that the haloes from the comprehensive sample with mass below the completeness limit constitute a biased sample, which are lying in a dense environment compared to observations. We only include central galaxies in the calculation as the haloes of satellites galaxies will have suffered tidal disruption. However, as the hydrodynamical simulations feature stars in the halo (which can be treated as ICL), we also include the mass of the ICL in the calculation for the SAMs \sag\ and \sage. Therefore, the central galaxy here is BCG+ICL. In agreement with our previous findings in Figs.~\ref{fig:image} and \ref{fig:bf}, \gadgetmusic\ has the highest stellar-to-halo-mass fraction. \sage, \sag\ and \gadgetx\ are in the second family, which tend to agree with the observational result at the lower mass end, but deviate from them at the massive end. \galacticus, which does not have ICL included, is in better agreement with \cite{Rodriguez-Puebla2017, Yang2009}. Moreover, we confirm that \sage\ also presents a better agreement with the observations if the ICL is excluded. We further note here that the BCG mass from \cite{Ragone-Figueroa2018} (a similar cluster simulation based on \gadgetx) is in a good agreement with observational results after applying a cut in radius. In addition, \cite{Pillepich2018} also reported that the exact functional form and magnitude of the stellar mass to halo mass relation strongly depend on the definition of a central galaxy's stellar mass. Therefore, the differences shown in this plot could be simply caused by the definition of the central galaxy. We further include the fitting result from \cite{Kravtsov2018}, who claim to account for the stellar mass in the same way as the model results here, i.e. BCG mass plus ICL mass. It is interesting to see that their $M_{\rm BCG}$ - $M_{\rm halo}$ relation is much closer to the results from our models (except \gadgetmusic\ which is far from any observation results and \galacticus\ which does not include ICL), especially at $M_{\rm halo} \lesssim 10^{14} \hMsun$. However, the offsets between the solid purple line and our model results (including \galacticus\ when compared with the observational results that do not include ICL) are still large for the most massive haloes. This means that the quenching of star formation in these massive clusters is still problematic for the models investigated here.

In order to check for the properties and influence of the ICL, for example the fraction, the evolution and the connection to the SHMR, we will perform a detailed investigation for both SAMs and the hydrodynamical simulations through carefully separating BCG from ICL, and present the results in a follow-up work (Ca\~nas et al. in prep.).

\subsubsection{Stellar mass function for satellite galaxies}
\begin{figure}
	\includegraphics[width=\linewidth]{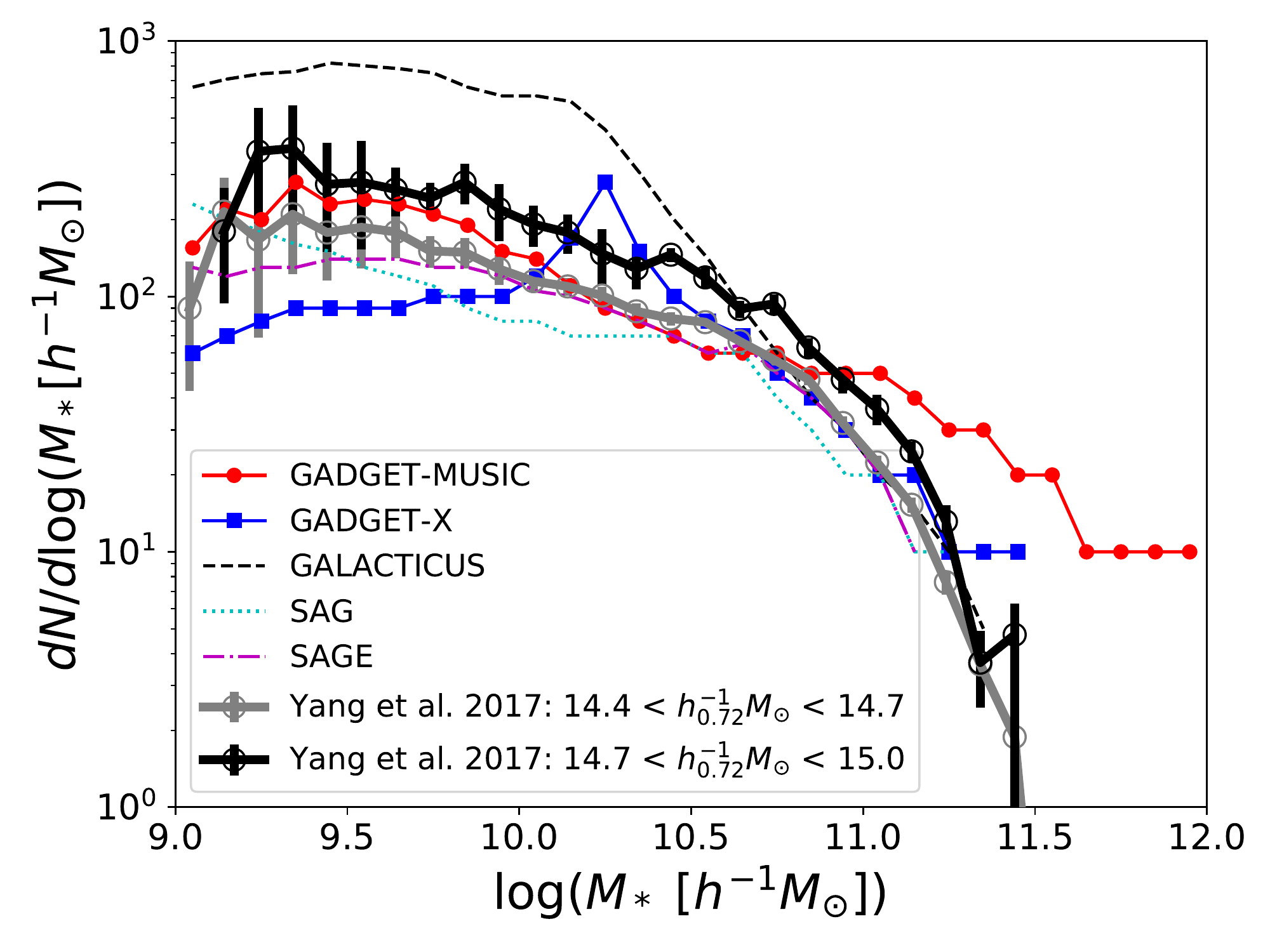}
    \caption{The median stellar mass function of satellite galaxies within the mass-complete cluster sample. \gadgetmusic\ is shown with a red line with circle symbols and \gadgetx\ with a blue line with square symbols. The three SAMs are presented by different lines: \galacticus\ as a black dashed line, \sag\ as a cyan dotted line and \sage\ as a magenta dot-dashed line. They are compared with observational results from \citet{Yang2018}, which are shown in thick black for halo mass range $[10^{14.7} - 10^{15} \ h^{-1}_{0.72}M_{\odot}]$ and thick grey for halo mass range $[10^{14.4} - 10^{14.7} \ h^{-1}_{0.72}M_{\odot}]$, both lines include error-bars.}
    \label{fig:ssmf}
\end{figure}

Though the satellite-galaxy stellar-mass function is not a scaling relation, we briefly switch focus from central galaxies to satellite galaxies and present the result in this subsection. We only use the mass-complete sample for this investigation and limit our satellite galaxies to objects within $R_{200}$ as per the observational sample. We show the stellar mass function -- median averaged over all clusters -- in Fig. \ref{fig:ssmf}. As indicated in the legend, different style thin lines represent different versions of the simulations and SAMs, while observational results from \cite{Yang2018} at two different cluster mass bins are highlighted as thick lines. Note that the complete cluster sample is used here without further binning in halo mass, because its mass limit is basically comparable with Yang's most massive mass bin. The lower mass bin from Yang's catalogue is presented here to aid the comparison. The horizontal extensions to the red and blue curves are artefacts of the median values. Compared to the observational results, \gadgetmusic\ has more massive satellite galaxies with masses $M_* > 10^{11.5} \hMsun$. \gadgetx\ shows a slightly reduced number of satellite galaxies towards the low mass end. \galacticus\ features the opposite trend. These deviations from the actual observations can be understood as an overabundance of massive satellite galaxies in \gadgetmusic\ due to the lack of AGN feedback; too few low mass satellite galaxies in \gadgetx\ can be caused by either a resolution issue (note that galaxies of $M_* \approx 10^{10} \hMsun$ only contain a few hundreds of stellar particles due to the poor simulation resolution) or the striped/heated gas due to the Wendland kernel and feedback; too many low mass satellite galaxies in \galacticus\ is because of a surplus of orphan galaxies \citep[see Table~2 in][]{Knebe2018}. \sag\ and \sage\ seem not to suffer from this problem due to their different treatment of the orphan galaxy population. We refer to \cite{Pujol2017} for a detailed comparison of the orphan galaxies between different SAMs. However, we note that the scatter across models seen here is at the level found in previous comparisons of theoretically modelled galaxy stellar mass functions of galaxies \citep[][]{Knebe2015,Knebe2018}. 

\subsubsection{Optical scaling relations}
\begin{figure}
	\includegraphics[width=\linewidth]{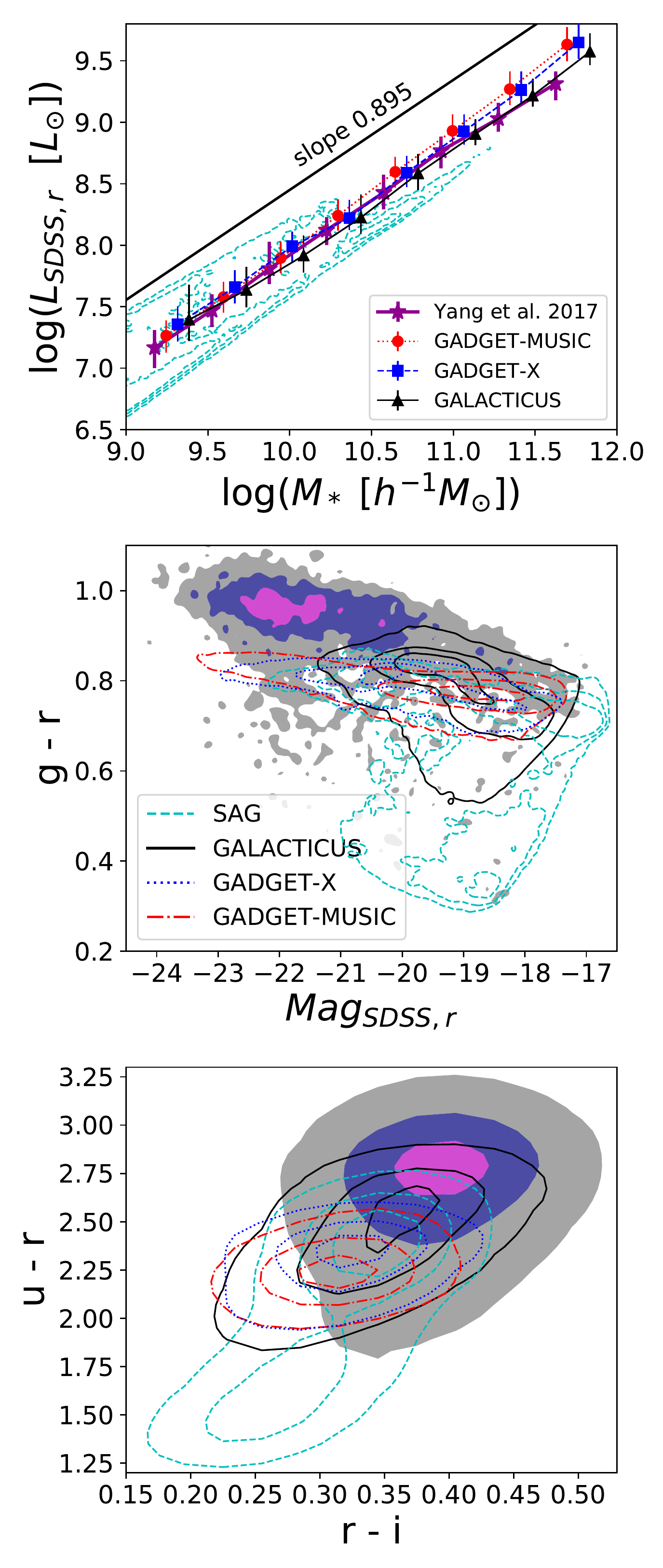}
    \vspace{-0.8cm}
    \caption{Top panel: the luminosity-stellar mass relation for all the galaxies inside the clusters (using the SDSS-$r$ band). As indicated in the legend, different symbols (median value) with error-bars ($16^{\rm th} - 84^{\rm th}$ percentile) are for different models and for the observational result from \citet{Yang2018}, while the result from \sag\ is presented in cyan contours. The top sloping black line (shifted up by 0.5 dex) shows the slope 0.895 which fits both the models and the observational result. Middle panel: the colour-magnitude relation for the galaxies inside the clusters. Bottom panel: the colour-colour relation for galaxies inside the clusters. The legend in the middle panel distinguishes the colours for the models with different line styles for both middle and bottom panels with the colour map is again from \citet{Yang2018}.}
    \label{fig:lc}
\end{figure}

We continue to investigate the correlations between luminosity/magnitude, stellar mass, and colours by comparing our modelled galaxies to the observational results from \cite{Yang2018}. We again only use the galaxies from our mass-complete sample here. For a fair comparison to our theoretical data, we apply the same mass cut ($M_{\rm 200} \geq 6.42 \times 10^{14} \hMsun$) to the group catalogue of \cite{Yang2018} and use all the satellites and central galaxies with $M_* > 10^9 \hMsun$ in these selected groups (the same criteria also applied to our complete sample). The results can be viewed in \Fig{fig:lc} where the top panel shows the luminosity-stellar mass relation (based upon the SDSS-$r$ band), the middle panel presents the $g-r$ colour -- magnitude (at SDSS-$r$ band) relation and the bottom panel shows the colour-colour relation with $u-r$ versus $r-i$. Note that the \sage\ model does not provide luminosities \textit{ab initio} and has hence been excluded from this plot. 
Similar to Fig.~\ref{fig:bf}, the contours are drawn at the same percentile density levels (16$^{\rm th}$, 50$^{\rm th}$ and 84$^{\rm th}$) after a normalized 2D binning with the observational results shown as different colour-filled areas. 

In the top panel, we recover a very tight correlation between luminosity and stellar mass with little variation between observation and the models (excluding \sag). \gadgetmusic, \gadgetx, \galacticus\ and Yang's observational results are binned only in stellar mass and presented by symbols with error-bars indicating the $16^{\rm th} - 84^{\rm th}$ percentile. While \sag, which tends to have a larger spread in luminosity, is shown with cyan contours. Moreover, we fit the $M_*$ - luminosity relation for the models (excluding \sag) and the observational result with a linear function $f(x) = a x$. We find that all the models give a consistent result with a slope of 0.895, which is shown by the solid black line shifted up by 0.5 dex in the top panel. In the colour-magnitude relation, both hydrodynamical simulations and SAMs show values $\sim 0.1 - 0.2$ below the $g-r$ colour of the observations. There are very few galaxies with a $g-r$ colour less than 0.7 in the observational results compared to the SAMs. This indicates that the SAMs -- as applied to a full cosmological simulation here -- fail to reduce their star-forming galaxies sufficiently in the cluster environment. The hydrodynamical simulations also have problems in ceasing star formation, especially for the brightest galaxies. For the colour-colour plot presented in the bottom panel, the results from the two hydrodynamical simulations are in agreement with the two SAMs. Although they all show a noticeable overlap with the observational results, the peaks for the four models are slightly shifted to smaller values in both colours compared to the observations.

\subsection{Gas scaling relations}

For the gas scaling relations, we now use our comprehensive sample of objects, but restrict our analysis to the hydrodynamical simulations for which we have immediate access to multiple gas properties. We confine the analysis to $M_{500}$ by reselecting all gas particles within $R_{500}$ to facilitate direct comparison to the observational results. 

\begin{figure}
	\includegraphics[width=\linewidth]{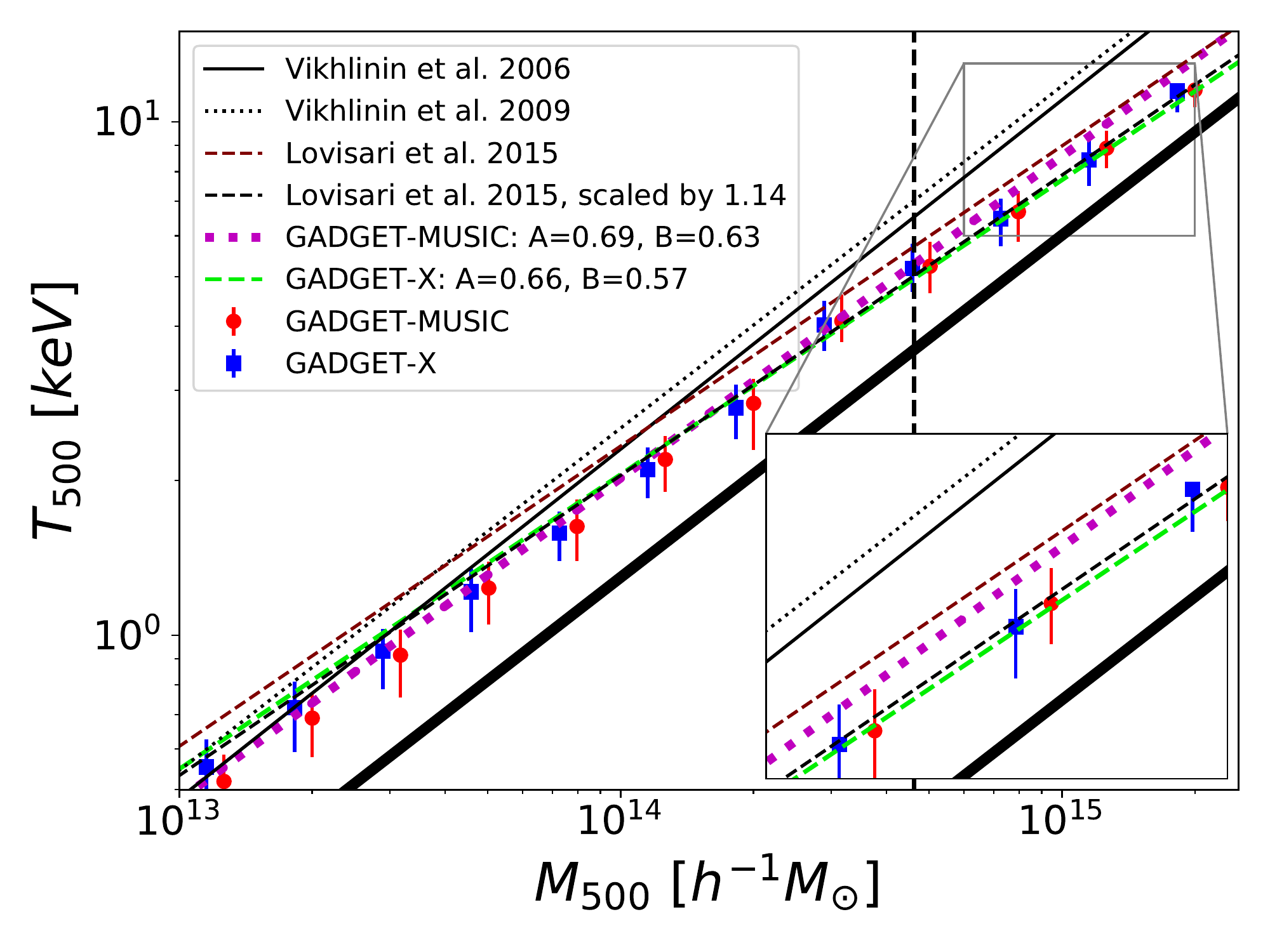}
    \caption{The temperature-mass relation for the clusters from the two hydrodynamical simulations. Red filled circles (blue filled squares) with error-bars ($16^{\rm th} - 84^{\rm th}$ percentile) are for \gadgetmusic\ (for \gadgetx). The solid and dotted black lines show the observational results from \citet{Vikhlinin2006} and \citet{Vikhlinin2009} respectively. The maroon dashed line shows the fitting result from \citet{Lovisari2015} (scaled by 1.14 as a black dashed line). Our fitting results from \gadgetmusic\ and \gadgetx\ are presented by magenta dotted and lime dashed lines respectively. The thick solid black line shows the self-similar relation $T_{500} \propto M_{500}^{2/3}$ predicted from non-radiative simulations.}
    \label{fig:tm}
\end{figure}
\begin{table}
	\centering
	\caption{The fitted parameters for the $T_{500} - M_{500}$ relation with fitting function: $T_{500} = 10^{\rm A} (M_{500}/6\times10^{14}\ {\rm M}_\odot)^{\rm B}$, see Eq.~\ref{eq:fit_tm} for details.}
	\label{tab:fittm}
	\begin{tabular}{lcc} 
		\hline
		Simulation & A & B \\
		\hline
		\gadgetmusic\	& 0.688$\pm 0.011$ & 0.627$\pm 0.007$\\
		\gadgetx\		& 0.663$\pm 0.012$ & 0.574$\pm 0.008$\\
		\hline
	\end{tabular}
\end{table}

We first investigate the temperature-mass ($T-M$) relation. The gas temperature is computed using the mass weighted temperature formula $T = \sum_{\rm i} T_{\rm i} m_{\rm i}/\sum_{\rm i} m_{\rm i}$, where $T_{\rm i}$ and $m_{\rm i}$ are the temperature and mass of a gas particle, respectively. In \Fig{fig:tm}, we show the relation between the mass-weighted gas temperature and $M_{\rm 500}$. 
We apply a simple linear fitting function in logarithm space to fit the data from all the samples: 
\begin{equation}
T_{500} = 10^{\rm A} \left(\frac{M_{500}}{6\times10^{14} \ {\rm M}_\odot} \right)^{\rm B}.
\label{eq:fit_tm}
\end{equation}
We especially note here that we exclude the $h$ in the normalization mass of the fitting equation (\ref{eq:fit_tm}). 

Since, as discussed above, our comprehensive cluster sample is not complete at the low mass end, data points below our completeness threshold are weighted according to their completeness during the fitting. As the comprehensive sample forms a mass-incomplete set of haloes they may conceivably be a biased dataset. Such a bias could in principle arise due to their physical proximity to a larger halo but how to accurately quantify such a bias, if it exists, is unclear. Best-fit curves are shown as a magenta dotted line for \gadgetmusic\ and a green dashed line for \gadgetx; the parameters are summarized both in the legend and Table~\ref{tab:fittm}. Since the low mass data has less weight and there are few clusters in the high mass range, it is not surprising to see that the fitting lines are offset from the symbols which show the median values in each mass bin.

The best fitting parameters are slightly different between the two hydrodynamical simulations: \gadgetmusic\ has a steeper slope close to the self-similar relation with B = 2/3 (\citealt{Kaiser1986}, also predicted by the non-radiative simulations, see \citealt{Bryan1998,Thomas2001} for example) compared to \gadgetx. This is mainly caused by the low temperature of the clusters with small halo mass. Compared to the results from \cite{Vikhlinin2006,Vikhlinin2009}, there is a good agreement at low halo mass with our simulations. However, there is a clear offset between our simulation result and their results for massive haloes. This could be caused by the hydrostatic method used in observations which can underestimate the total mass due to a non-thermal pressure component. This bias has been corrected in \cite{Lovisari2015}, which, although it is still above our best fit lines, is closer to our data for the most massive haloes (closer to \gadgetmusic\ than to \gadgetx). In addition, their result is also slightly higher than our simulation results at low halo mass. This is because of the spectrum-weighted temperature adopted in \cite{Lovisari2015}, which is about 14 per cent higher than the mass weighted temperature \citep{Vikhlinin2006, MUSICII}. We follow \cite{MUSICII} by correcting for this difference by scaling down the fitting function from \cite{Lovisari2015} by a factor of 1.14 (black dashed line in \Fig{fig:tm}). This produces a very good match to the fitting result from \gadgetx. It is worth noting that the self-similar relation does not provide a good fit to our data \citep[see also][]{Truong2018}. Lastly, \cite{Truong2018} reported lower temperatures than observed resulting in a normalization shift of about 10 per cent for the $T-M$ relation for their AGN model. Similarly, \cite{Henden2018} also found such a difference with zoomed-in cluster simulations. However, they claimed this is most likely caused by the underestimated total mass due to the biased X-ray hydrostatic mass than a lower temperature in their simulation.

\bigskip

\begin{figure}
	\includegraphics[width=\linewidth]{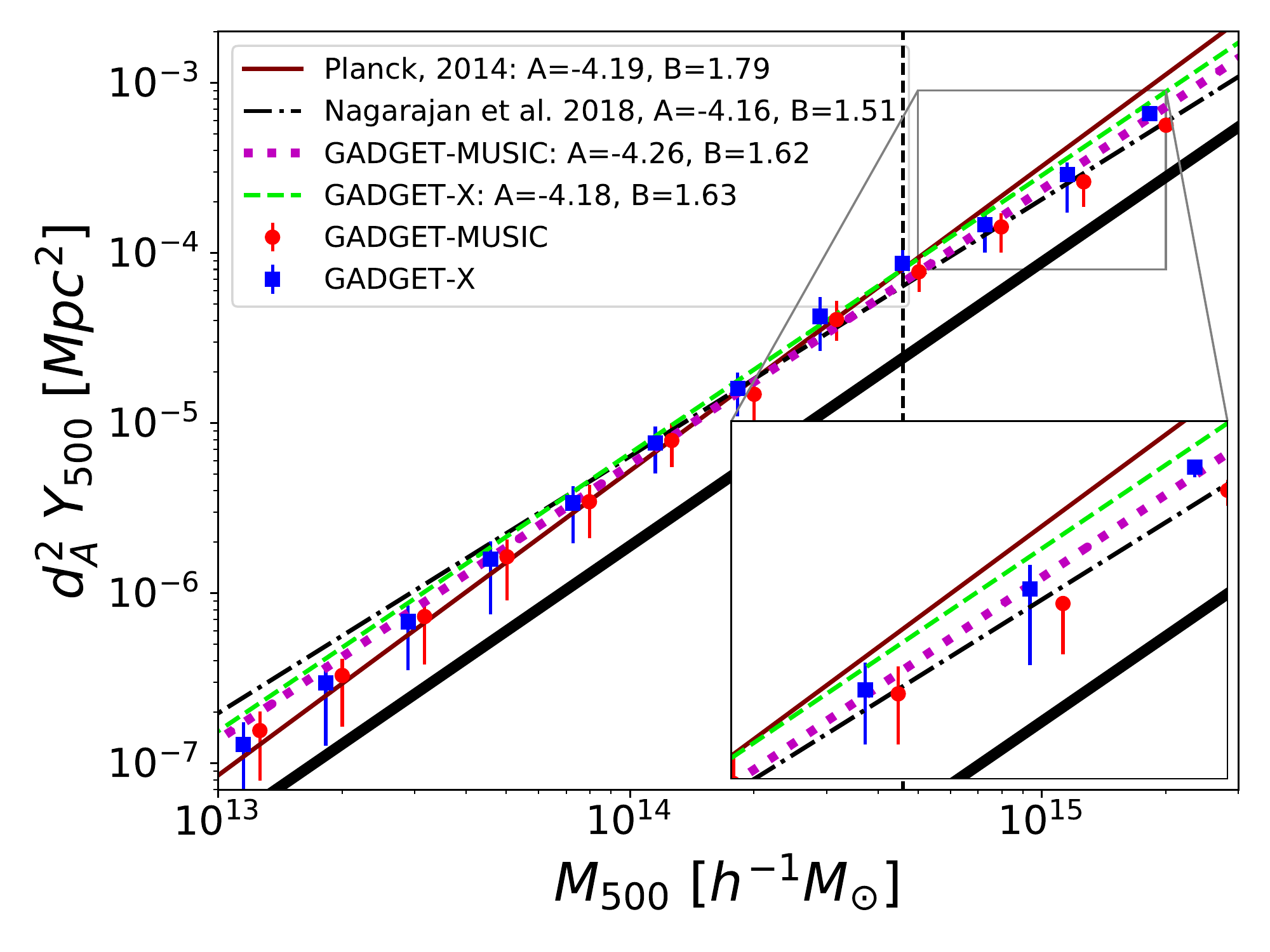}
    \caption{The $Y_{500} - M_{500}$ relation. Similar to Fig.~\ref{fig:tm}, red circles (median value) with error-bars ($16^{\rm th} - 84^{\rm th}$ percentile) are for \gadgetmusic\ while blue squares with error-bars are for \gadgetx. The thin maroon line comes from the {\it Planck} observation \citep{Planck2014} and the dash-dotted line is the fitted result from \citet{Nagarajan2018} with cluster mass estimated by the weak-lensing method. While the black dotted and lime dashed lines show our fitting results for \gadgetmusic\ and \gadgetx\ respectively. The lower thick black line shows the self-similar relation $Y_{500} \propto M_{500}^{5/3}$.}
    \label{fig:ym}
\end{figure}

\begin{table}
	\centering
	\caption{The fitted parameters for the $Y_{500} - M_{500}$ relation. See equation \ref{eq:fitym} for details.}
	\label{tab:fitym}
	\begin{tabular}{lcc} 
		\hline
		Simulation & A & B \\
		\hline
		\gadgetmusic\	& -4.26$\pm 0.07$ & 1.62$\pm 0.31$\\
		\gadgetx\		& -4.18$\pm 0.07$ & 1.63$\pm 0.29$\\
		\hline
	\end{tabular}
\end{table}

The Sunyaev-Zel'dovich (SZ) effect \citep{Sunyaev1970} -- which is the diffusion of cosmic microwave background photons within a hot plasma (normally inside galaxy clusters) due to inverse Compton scattering -- provides a unique view of a galaxy cluster. Therefore, it has become one of the most powerful cosmological tools used to study the ICM, as well as the nature of the dark matter and dark energy components of the Universe. Numerous works have been devoted to investigate and understand this effect, both observationally \citep[e.g.][]{Staniszewski2009,Marriage2011,Planck2015} and theoretically by means of cosmological simulations \citep[e.g.][]{daSilva2000,MUSICI,LeBrun2015,Dolag2016}.


The thermal SZ signal is characterized by the dimensionless Compton $y$-parameter, which is defined as
\begin{equation}
y = \frac{\sigma_T k_{\rm B}}{m_{\rm e} c^2} \int n_{\rm e} T_{\rm e} dl,
\end{equation}
here $\sigma_{\rm T}$ is the Thomson cross-section, $k_{\rm B}$ the Boltzmann constant, $c$ the speed of light, $m_{\rm e}$ the electron rest-mass, $n_{\rm e}$ the electron number density and $T_{\rm e}$ the electron temperature. The integration is done along the observer's line of sight. In the hydrodynamical simulations, the electron number density, $n_{\rm e}$, for one gas particle can be represented as $n_{\rm e}$ = $N_{\rm e}/dV$ = $N_{\rm e}/dA/dl$, here $N_{\rm e}$ is the number of electrons in the gas particle, $dV$ is its spatial volume which is broken down into $dA$ (the projected area) and $dl$ (the line of sight distance). Therefore, the integration can be represented by the summation \citep{MUSICI,LeBrun2015}:
\begin{equation}
y = \frac{\sigma_T k_{\rm B}}{m_{\rm e} c^2 dA} \sum_{\rm i} T_{\rm i} N_{\rm e,i} W(r,h_{\rm i}),
\end{equation}
here we applied the same SPH smoothing kernel $W(r,h_{\rm i})$ as the hydrodynamical simulation to smear the $y$ signal from each gas particle to the projected image pixels where $h_{\rm i}$ is the gas smoothing length from the simulations. It is worth noting that the number of electrons per gas particle is metallicity dependent: $N_{\rm i} = \frac{N_{\rm e} m_{\rm i} (1 - Z - Y_{\rm He})}{\mu m_{\rm p}}$, where $N_{\rm e}$ is the number of ionized electrons per hydrogen particle, $m_{\rm i}$ the mass of the gas particle, $Z$ the metallicity of the gas particle, $Y_{\rm He}$ the helium mass fraction of the gas particle, $\mu$ the mean molecular weight and $m_{\rm p}$ the proton mass.\footnote{The analysis pipeline for this calculation is publicly available as a python package from \url{https://github.com/weiguangcui/pymsz}.} 

The integrated Comptonization parameter $Y$ over an aperture inside $R$ is given by: 
\begin{equation}
Y = \int y d\Omega = \sum_{\rm i}^{{\rm i}\ \in\ R} y_{\rm i},
\end{equation}
where $\Omega$ is a solid angle, which can be expressed as an aperture of radius $R$. In observations, this $Y$ parameter is normally re-expressed as $d_A(z)^2 E(z) Y$, where $d_A(z)$ is the angular diameter distance and $E(z) = H(z)/H0 = \sqrt[]{\Omega_{\rm m} (1+z)^3 + \Omega_{\rm \Lambda}}$ gives the redshift evolution of the Hubble parameter, $H(z)$, in a flat $\Lambda$CDM Universe. Here we are only presenting clusters at redshift $z=0$, for which $E(z) = 1$. In the subsequent analysis, we focus on $Y_{500}$ within an aperture of $R_{500}$. Moreover, we only present projected results in the x-y plane here. Since we have a large number of samples, the projection effect should have a negligible impact on our results.

In \Fig{fig:ym}, we show the scaling between $Y_{500}$ and $M_{\rm 500}$. Similar to \Fig{fig:tm}, symbols with error-bars are calculated from our comprehensive sample by binning in mass. We refer to the legend in \Fig{fig:ym} for further details. Here, we adopt a similar functional form as used for the $T-M$ relation to fit the data from our comprehensive sample:
\begin{equation}
d_A^2 Y_{500} = 10^{\rm A} \left(\frac{M_{500}}{6 \times 10^{14}\ {\rm M}_\odot} \right)^{\rm B}.
\label{eq:fitym}
\end{equation}
The best-fitting parameters from \cite{Planck2014} are A = -4.19 and B = 1.79, which relies on mass estimates from a mass-proxy relation due to \cite{Kravtsov2006}. The fitting result from \cite{Nagarajan2018} which used the weak lensing mass of the APEX-SZ clusters, is shown as a purple dash-dotted line with A = -4.16 and B = 1.51. We fit our simulation data to the same function and present the results in \Fig{fig:ym} for \gadgetmusic\ as a black dotted line and for \gadgetx\ as a green dashed line. The value of the best-fitting parameters are shown in both the figure legend and Table~\ref{tab:fitym}. Compared to the best-fit Planck relation, our simulation results have a slightly flatter slope. However, comparing to the result from \cite{Nagarajan2018} who used a more precise mass estimation method, both \gadgetx\ and \gadgetmusic\ are slightly above (similar offsets as comparing with the Planck result) the purple line at the high mass end. On the contrary, the Planck (APEX-SZ) fitting line is under (above) the simulation results at the low mass end ($M_{500} < 10^{13.5} \hMsun$). In addition, \gadgetx\ only shows a marginally higher amplitude than \gadgetmusic, especially at the high-mass end of the relation. Both are also in agreement with the self-similar relation with B = 5/3 \citep[e.g.][]{Bonamente2008}. This means that the scaling between $M_{500}$ and $Y_{500}$ is almost independent of the gas physics and is the more robust relation, which is in agreement with \citet{Planelles2017,Truong2018}, for example. It is worth noting that neither observations used mass $M_{500} < 10^{14} \hMsun$ to do the fitting. It is interesting to see that this scaling relation extends down to mass $M_{500} = 10^{13} \hMsun$ for our models.

\section{Conclusions} \label{sec:conclusion}
In this paper we introduce {\sc The Three Hundred} project, i.e. a data base of more than 300 synthetic galaxy clusters with mass $M_{\rm 200}>6\times 10^{14}\hMsun$. The clusters have been individually modelled in a cosmological volume of side length $1\Gpc$ with all the relevant baryonic physics (including AGN feedback) using the `modern' SPH code \textsc{Gadget-X} \citep{Beck2016}. The large re-simulation regions of radius $15\Mpc$ -- centred on the 324 most massive galaxy clusters as found in the parent dark matter only MDPL2 simulation -- contain many additional objects, in total about $ 5500$ objects with a mass $M_{\rm 200}>10^{13}\hMsun$. This suite of massive galaxy clusters therefore not only allows to study the formation and evolution of a mass-complete sample, but also carefully investigate their environments and the preprocessing of material entering the galaxy cluster. 

This introductory paper focuses on presenting the galaxy clusters by primarily studying their redshift $z=0$ properties and comparing them to observational data. This serves as a validation of the public data. Additionally, we do have at our disposal the same suite of clusters, but simulated with a `classical' SPH technique and without AGN feedback \citep[i.e. the \textsc{Gadget-MUSIC} code,][]{MUSICI}. This forms a comparison benchmark, demonstrating the differences that choices surrounding physical prescriptions can make. We further presented -- where appropriate -- the results as obtained via three distinct SAMs (\galacticus, \sag, and \sage) that were applied to the underlying dark matter only MDPL2 simulation. A comparison between full physics simulations and semi-analytic models of galaxy formation on this scale or with this number of objects adds to existing efforts of gauging the relevance of various physical processes and its numerical modelling. In subsequent papers we will apply a more elaborate analysis including redshift evolution and formation processes.

We find that our clusters are in reasonable agreement with observations and summarize our main findings as follows:
\begin{itemize}

\item The cluster mass difference between the hydrodynamical simulations and their dark-matter-only counterpart is very small for $M_{200}$, with about 5 per cent scatter. However, $M_{500}$ is about 2-6 percent higher in the hydrodynamical simulation than their MDPL2 counterparts at $4\times 10^{14} \lsim M_{500} \lsim 10^{15}$, with a large scatter of about 10 per cent. Using the dynamically relaxed sample reduces the scatter in half, but does not change the systematic differences.

\item The dynamically relaxed cluster sample has a $c-M$ relation which appears to be flat for \gadgetx\ across the considered mass range. The concentrations for \gadgetmusic\ are generally larger (factor of approx. 1.3) and in better agreement with observations. In both models the concentrations of the hydrodynamically modelled clusters are larger than those of their dark matter only counterparts; for \gadgetmusic\ this applies to the full mass range whereas for \gadgetx\ concentrations appear unaffected by the inclusion of baryon physics beyond $10^{15}\hMsun$.

\item \gadgetx\ shows baryonic fractions at $M_{500} \gsim 10^{14} \hMsun$ that are generally in agreement with observations, while \gadgetmusic\ forms too many stars due to the lack of AGN feedback. \sag\ has the highest gas fraction and the lowest stellar fraction in haloes. \sage\ and \galacticus\ share similar gas fractions and stellar fractions (slightly higher in \sage\ than \galacticus).

\item Besides \galacticus, all the models included in this study do not produce a stellar halo mass relation that is consistent with observations. This could be caused by the inclusion of the ICL. Even comparing with the observational result from \cite{Kravtsov2018}, which has ICL included, the BCGs in our modelled clusters ($M_{\rm halo} \gsim 10^{14.5}$) are still massive.

\item For the stellar mass function of the satellite galaxies, \gadgetmusic\ over produces the number of massive satellites. At lower stellar mass, \galacticus\ (\gadgetx) has more (less) satellites than the observations. 

\item The hydro runs and \galacticus\ show a linear (with a slope of 0.895) luminosity-mass relation which is very consistent with the observational result. All the models fail to represent the peak position from observations for the colour-magnitude and colour-colour contour.

\item For the gas scaling relations, both \gadgetx\ and \gadgetmusic\ are generally in agreement with the observational temperature-mass and $Y_{500}$-mass relations. The fitting for the hydrodynamical simulations extends to $10^{13} \hMsun$, which shows the power of the scaling relation. The small difference between the two simulations indicates that baryonic processes only have a weak influence on these relations \citep[see also][]{Hahn2017}.
\end{itemize}

In addition to the publication of the simulations and halo catalogues, we plan to make publicly available a multi-wavelength mock observation database (Cui et al., in prep.) which will include observational mock images from radio/SZ, optical bands to X-rays of all our simulated clusters at different redshifts. We will also provide gravitational lensing images and investigate the lensing efficiency in a follow-up paper (Vega-Ferrero et al. in prep.).\\

We close with the concluding remark that our theoretically modelled galaxies and galaxy clusters generally present similar results and matches to observations - at least on certain scales of interest. However, we do see deviations in multiple aspects between these models and the observations, especially for the massive central galaxy (BCG+ICL). To understand the disagreements and to connect them with the input sub-grid baryonic models, we need to a) extend the comparisons to even smaller scales than the ones presented here, b) consistently derive quantities by mimicking observations more quantitatively, and c) track the impact of these baryonic models over a wider range of redshifts. Eventually, as our cluster sample contains different physical implementations of various baryonic processes from both hydrodynamic and SAM modelling, this will allow us to investigate, understand, and pin down the differences between our results and connect them back with the underlying physics. Several such follow-up works are already under way and will be presented separately from this introductory paper (e.g., Vega-Mart\'inez et al. in prep.; Li et al. in prep.). Further, in a companion paper \citep{Mostoghiu2018}, we investigate the density profile of these clusters together with its evolution. And in \citep{Wang2018} the analysis is extended to the comprehensive sample of haloes in the re-simulation regions, investigating how the environment affects their properties and, in particular, the star formation rate. Furthermore, disentangling the BCG from ICL (Ca\~nas et al., in prep.) will help us to understand the too massive central galaxy problem in detail.

\section*{Acknowledgements}
The work has received financial support from the European Union's Horizon 2020 Research and Innovation programme under the Marie Sklodowskaw-Curie grant agreement number 734374, i.e. the LACEGAL project\footnote{https://cordis.europa.eu/project/rcn/207630\_en.html}. The workshop where this work has been finished was sponsored in part by the Higgs Centre for Theoretical Physics at the University of Edinburgh.

The authors would like to thank The Red Espa\~nola de Supercomputaci\'on for granting us computing time at the MareNostrum Supercomputer of the BSC-CNS where most of the cluster simulations have been performed. The MDPL2 simulation has been performed at LRZ Munich within the project pr87yi. The CosmoSim database (\href{https://www.cosmosim.org}{https://www.cosmosim.org}) is a service by the Leibniz Institute for Astrophysics Potsdam (AIP). Part of the computations with \gadgetx\ have also been performed at the `Leibniz-Rechenzentrum' with CPU time assigned to the Project `pr83li'.

This work has made extensive use of the {\sc Python} packages --- {\sc Ipython} with its Jupyter notebook \citep{ipython}, {\sc NumPy} \citep{NumPy} and {\sc SciPy} \citep{Scipya,Scipyb}. All the figures in this paper are plotted using the python {\sc matplotlib} package \citep{Matplotlib}. This research has made use of NASA's Astrophysics Data System and the arXiv preprint server. 

\begin{CJK}{UTF8}{gbsn} 
WC, AK, GY and RM are supported by the {\it Ministerio de Econom\'ia y Competitividad} and the {\it Fondo Europeo de Desarrollo Regional} (MINECO/FEDER, UE) in Spain through grant AYA2015-63810-P. WC further thanks TaiLai Cui (崔泰莱) for all the joys. AK is also supported by the Spanish Red Consolider MultiDark FPA2017‐90566‐REDC and further thanks Krog for making the days counts.
CP acknowledges the Australia Research Council (ARC) Centre of Excellence (CoE) ASTRO 3D through project number CE170100013.
PJE is supported by the ARC CoE ASTRO 3D through project number CE170100013.
SB acknowledged financial support from PRIN-MIUR grant 2015W7KAWC, the agreement ASI-INAF n.2017-14-H.0, the INFN INDARK grant, the EU H2020 Research and Innovation Programme under the ExaNeSt project (Grant Agreement No. 671553).
ER acknowledge the ExaNeSt and Euro Exa projects, funded by the European Union's Horizon 2020 research and innovation programme under grant agreement No 671553 and No 754337 and financial contribution from the agreement ASI-INAF n.2017-14-H.0.
DS' fellowship is funded by the \textit{Spanish Ministry of Economy and Competitiveness} (MINECO) under the 2014 \textit{Severo Ochoa} Predoctoral Training Programme.
J.V-F acknowledges the hospitality of the Physics \& Astronomy Department at the University of Pennsylvania for hosting him during the preparation of this work.
YW is supported by the national science foundation of China (No. 11643005).
XY is supported by the National Key Basic Research Program of China (No. 2015CB857002), national science foundation of China (No. 11233005, 11621303).
JTA acknowledges support from a postgraduate award from STFC.
SAC acknowledges funding from {\it Consejo Nacional de Investigaciones Cient\'{\i}ficas y T\'ecnicas} (CONICET, PIP-0387), {\it Agencia Nacional de Promoci\'on Cient\'ifica y Tecnol\'ogica} (ANPCyT, PICT-2013-0317), and {\it Universidad Nacional de La Plata} (G11-124), Argentina.
CVM acknowledges CONICET, Argentina, for their supporting fellowships.
ASB, GC and MDP are supported by Sapienza University of Rome-Progetti di Ricerca Anno 2016. ASB also acknowledges funding from Sapienza Universit\`a di Roma under minor grant Progetti per Avvio alla Ricerca Anno 2017, prot. AR11715C82402BC7.
RC is supported by the MERAC foundation postdoctoral grant awarded to Claudia Lagos and by the Consejo Nacional de Ciencia y Tecnolog\'ia CONACYT CVU 520137 Scholar 290609 Overseas Scholarship 438594.
SE acknowledges financial contribution from the contracts NARO15 ASI-INAF I/037/12/0, ASI 2015-046-R.0 and ASI-INAF n.2017-14-H.0.
SEN is member of the Carrera del Investigador Cient\'{\i}fico of CONICET.
SP is supported by the Fundamental Research Program of Presidium of the RAS \#28.
JS acknowledges support from the ``Centre National d'\'etudes spatiales" (CNES) postdoctoral fellowship program as well as from the ``l'Or\'eal-UNESCO pour les femmes et la Science" fellowship program.
\end{CJK}


\bibliographystyle{mnras}
\bibliography{bib}


\appendix

\section{Evolution of the Halo Mass Function} \label{sec:hmf}
\begin{figure*}
	\includegraphics[width=\linewidth]{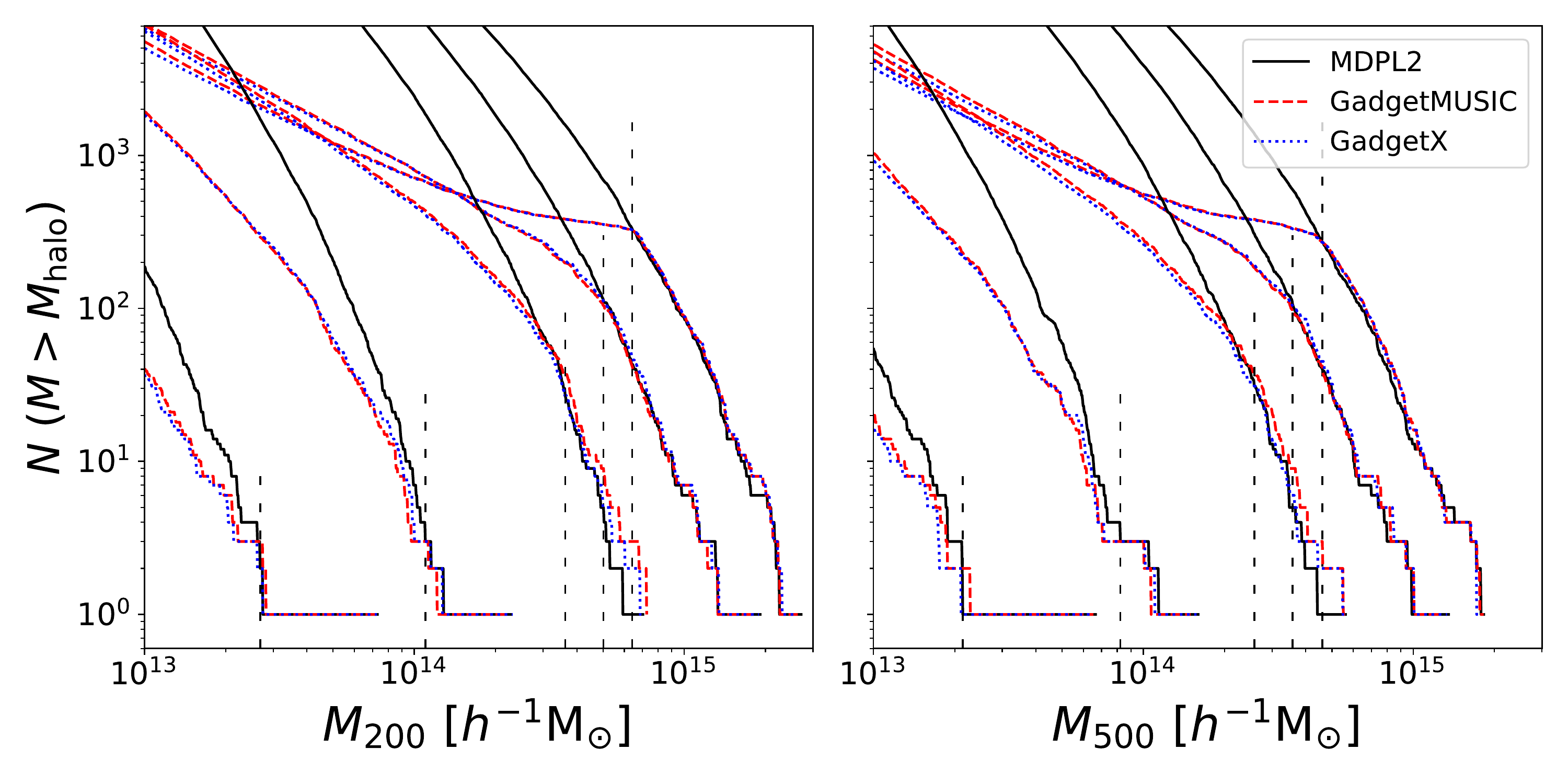}
    \caption{The cumulative halo mass function from different simulation runs for $M_{200}$ on the left-hand panel and $M_{500}$ on the right-hand panel. Different colour and line styles represent different simulations: solid black lines are for the DM-only MDPL2; red dashed lines are for \gadgetmusic\ and blue dotted lines are for \gadgetx. From left to right, we show the halo mass function at redshifts; $z = 4.0, 2.3, 1.0, 0.5, 0.0$ respectively. The dashed vertical lines indicate the mass to which we are complete (i.e. our simulation dataset contains all the haloes above this mass in the full simulation volume). Table \ref{tab:masscomplete} lists the exact values.}
    \label{fig:hmf}
\end{figure*}

\begin{table}
	\centering
	\caption{The mass-complete sample of the Three Hundred cluster catalogues at different redshifts. The first column shows the redshift. The second is the $M_{200}$ mass limit and the third column gives the values for $M_{500}$.}
	\label{tab:masscomplete}
	\begin{tabular}{lcc} 
		\hline
		redshift & $M_{200}$  & $M_{500}$ \\
        & [$10^{14} \hMsun$]& [$10^{14} \hMsun$]\\
		\hline
		0.0		& 6.42  & 4.60 \\
		0.5		& 5.02  & 3.57 \\
		1.0 	& 3.62  & 2.57 \\
        2.3 	& 1.10  & 0.82 \\
        4.0 	& 0.27  & 0.21 \\
		\hline
	\end{tabular}
\end{table}

Our 30 $\hMpc$ diameter re-simulated regions contain many more objects in addition to the central clusters. While there are lots of haloes in the region that surrounds the central cluster there would be many, many more similar haloes in the full volume. It is therefore important to understand the completeness of our comprehensive sample. Here, completeness refers to the total number of haloes above a given mass within a certain cosmological volume. The mass-complete sample in our hydrodynamic simulations is given by $N_{\rm hydro}(>M_{\rm X}) \geq N_{\rm MDPL2}(>M_{\rm X})$, here N is the total number of haloes above a certain mass $M_{\rm X}$ with X is the chosen mass overdensity e.g. $200$. i.e. this is the mass above which our sample contains every cluster in the full volume. Below this mass some haloes have not been captured by our re-simulation procedure.

In Fig.~\ref{fig:hmf}, we show the cumulative halo mass functions for the two mass definitions $M_{200}$ (left panel) and $M_{500}$ (right panel) as derived from MDPL2 (solid black lines), \gadgetmusic\ (red dashed lines) and \gadgetx\ (blue dot-dash lines). There are five families of lines inside each panel, which, from left to right, show the results at $z=4.0, 2.3, 1.0, 0.5, 0.0$. The mass function of the full halo catalogue from the MDPL2 is used here as a reference line. The vertical dashed lines indicates the mass down to which our sample is complete, determined by the crossing point between the \gadgetx\ and MDPL2 lines. The mass limit will slightly decrease at some redshifts if \gadgetmusic\ were to be used instead of \gadgetx. This is caused by the baryon effects, as \gadgetmusic\ forms more stars. In order to make sure the complete sample is chosen to be conservative, we use \gadgetx\ which returns a higher mass limit. We especially note here that the complete sample is based on the MDPL2 halo mass function. This matching ignores any baryon effects on the halo mass function. However, this could only affect a small number of them near the mass limitation (see Fig.~\ref{fig:hmd} for the mass difference). The precise values for these limits for our mass-complete sample are presented in table \ref{tab:masscomplete}. 

Below the mass-complete limits the completeness fraction, which will be used later to weight the fitting of the scaling relations, is calculated by the ratio of these lines. It is interesting to note that even at $z=1$ the number of clusters in the complete sample has fallen dramatically. This is because there is significant shuffling in the rank order of the most massive objects in the sample. The set of the largest objects at $z=4$ bears little relation to the largest objects at $z=0$ and one set does not evolve uniquely into the other. Conversely, the largest objects identified at $z=0$ are not all the largest objects at higher redshift and modelling them alone does not produce a large mass-complete sample at earlier times. We further note here that there is only a few mass-complete clusters at $z \geq 2.3$. The mass limits are more useful for indicating the boundary of the un-complete sample than for selecting the complete sample for statistical studies.

\section*{Affiliations}
\noindent
{\it
$^1$Departamento de F\'isica Te\'{o}rica, M\'{o}dulo 15 Universidad Aut\'{o}noma de Madrid, 28049 Madrid, Spain\\
$^{2}$Centro de Investigaci\'{o}n Avanzada en F\'{\i}sica Fundamental (CIAFF), Universidad Aut\'{o}noma de Madrid, 28049 Madrid, Spain \\
$^{3}$International Centre for Radio Astronomy Research, The University of Western Australia, 35 Stirling Highway, Crawley, \\ Western Australia 6009, Australia\\
$^{4}$School of Physics \& Astronomy, University of Nottingham, Nottingham NG7 2RD, UK\\
$^{5}$Institute for Astronomy, School of Physics \& Astronomy, The University of Edinburgh, Royal Observatory, Edinburgh EH9 3HJ, UK \\
$^{6}$University Observatory Munich, Scheinerstra{\ss}e 1, 81679 Munich, Germany\\
$^{7}$Max-Planck-Institute for Extraterrestrial Physics, Giessenbachstrasse 1, 85748 Garching, Germany\\
$^{8}$Dipartimento di Fisica, Sezione di Astronomia, Universit\'{a} di Trieste, via Tiepolo 11, I-34143 Trieste, Italy\\
$^{9}$INAF - Osservatorio Astronomico di Trieste, via Tiepolo 11, I-34143 Trieste, Italy\\
$^{10}$INFN - Sezione di Trieste, via Valerio 2, I-34127 Trieste, Italy\\
$^{11}$Max-Planck Institute for Astrophysics, Karl-Schwarzschild-Strabetae 1, D-85741 Garching, Germany \\
$^{12}$Instituto de F\'{i}sica Te\'{o}rica, (UAM/CSIC), Universidad Aut\'{o}noma de Madrid, Cantoblanco, E-28049 Madrid, Spain\\
$^{13}$IFCA, Instituto de F\'isica de Cantabria (UC-CSIC), Av. de Los Castros s/n, 39005 Santander, Spain\\
$^{14}$School of Physics and Astronomy, Sun Yat-sen University, 519082, Zhuhai, China\\
$^{15}$Department of Astronomy, Shanghai Key Laboratory for Particle Physics and Cosmology, Shanghai Jiao Tong University, Shanghai 200240, China\\
$^{16}$IFSA Collaborative Innovation Center, and Tsung-Dao Lee Institute, Shanghai Jiao Tong University, Shanghai 200240, China\\
$^{17}$Carnegie Observatories, 813 Santa Barbara Street, Pasadena, CA 91101, USA\\
$^{18}$Instituto de Astrof\'{i}sica de La Plata (CCT La Plata, CONICET, UNLP), Paseo del Bosque s/n, B1900FWA, La Plata, Argentina\\
$^{19}$Facultad de Ciencias Astron\'{o}micas y Geof\'{i}sicas, Universidad Nacional de La Plata, Paseo del Bosque s/n, B1900FWA, La Plata, Argentina\\
$^{20}$Centre for Astrophysics and Supercomputing, Swinburne University of Technology, Hawthorn, Victoria 3122, Australia\\
$^{21}$Department of Physics, Sapienza Universit\`{a} di Roma, p.le Aldo Moro 5, I-00185 Rome, Italy\\
$^{22}$Dipartimento di Fisica, Universit\`{a} di Roma Tor Vergata, via della Ricerca Scientifica 1, I-00133 Roma, Italy\\
$^{23}$South African Astronomical Observatory, PO Box 9, Observatory, Cape Town 7935, South Africa\\
$^{24}$Department of Physics and Astronomy, University of the Western Cape, Cape Town 7535, South Africa\\
$^{25}$INFN - Sezione di Roma, P.le A. Moro 2, I-00185 Roma, Italy \\
$^{26}$Dipartimento di Fisica - Universit\`{a} degli Studi di Torino, Via Pietro Giuria, 1, 10125 Torino, Italy\\
$^{27}$INAF, Osservatorio di Astrofisica e Scienza dello Spazio, via Pietro Gobetti 93/3, 40129 Bologna, Italy\\
$^{28}$INFN, Sezione di Bologna, viale Berti Pichat 6/2, I-40127 Bologna, Italy\\
$^{29}$Leibniz-Institut f\"{u}r Astrophysik, 14482 Potsdam, Germany\\
$^{30}$Instituto de Astronom\'{\i}a y F\'{\i}sica del Espacio (IAFE,
CONICET-UBA), CC 67, Suc. 28, 1428 Buenos Aires, Argentina\\
$^{31}$Facultad de Ciencias Exactas y Naturales (FCEyN), Universidad de Buenos Aires (UBA), Buenos Aires, Argentina\\
$^{32}$Department of Astronomy $\&$ Astrophysics, University of Toronto, Toronto, Canada\\
$^{33}$Astro Space Centre of Lebedev Physical Institute, Russian Academy of Sciences, Profsoyuznaja 84/32, 117997 Moscow, Russia\\
$^{34}$Univ Lyon, Univ Lyon1, Ens de Lyon, CNRS, Centre de Recherche Astrophysique de Lyon UMR5574, F-69230, Saint-Genis-Laval, France\\
$^{35}$CNRS and UPMC Univ. Paris 06, UMR 7095, Institut d'Astrophysique de Paris, 98 bis Boulevard Arago, F-75014 Paris, France\\
}

\bsp	
\label{lastpage}
\end{document}